%
%

\ifx\mnmacrosloaded\undefined 
%
%
%
%

\catcode `\@=11 

\def\@version{1.4}
\def\@verdate{22nd Feb 1994}

%
%
%
%


\newif\ifprod@font

\ifx\@typeface\undefined
  \def\@typeface{Comp. Modern}\prod@fontfalse
\else
  \prod@fonttrue 
\fi

\def\newfam{\alloc@8\fam\chardef\sixt@@n} 

\ifprod@font
\font\fiverm=mtr10 at 5pt
\font\fivebf=mtbx10 at 5pt
\font\fiveit=mtti10 at 5pt
\font\fivesl=mtsl10 at 5pt
\font\fivett=mttt10 at 5pt     \hyphenchar\fivett=-1
\font\fivecsc=mtcsc10 at 5pt
\font\fivesf=mtss10 at 5pt
\font\fivei=mtmi10 at 5pt      \skewchar\fivei='177
\font\fivemib=mtmib10 at 5pt   \skewchar\fivemib='177
\font\fivesy=mtsy10 at 5pt     \skewchar\fivesy='60
\font\fivesyb=mtbsy10 at 5pt   \skewchar\fivesyb='60

\font\sixrm=mtr10 at 6pt
\font\sixbf=mtbx10 at 6pt
\font\sixit=mtti10 at 6pt
\font\sixsl=mtsl10 at 6pt
\font\sixtt=mttt10 at 6pt      \hyphenchar\sixtt=-1
\font\sixcsc=mtcsc10 at 6pt
\font\sixsf=mtss10 at 6pt
\font\sixi=mtmi10 at 6pt       \skewchar\sixi='177
\font\sixmib=mtmib10 at 6pt    \skewchar\sixmib='177
\font\sixsy=mtsy10 at 6pt      \skewchar\sixsy='60
\font\sixsyb=mtbsy10 at 6pt    \skewchar\sixsyb='60

\font\sevenrm=mtr10 at 7pt
\font\sevenbf=mtbx10 at 7pt
\font\sevenit=mtti10 at 7pt
\font\sevensl=mtsl10 at 7pt
\font\seventt=mttt10 at 7pt     \hyphenchar\seventt=-1
\font\sevencsc=mtcsc10 at 7pt
\font\sevensf=mtss10 at 7pt
\font\seveni=mtmi10 at 7pt      \skewchar\seveni='177
\font\sevenmib=mtmib10 at 7pt   \skewchar\sevenmib='177
\font\sevensy=mtsy10 at 7pt     \skewchar\sevensy='60
\font\sevensyb=mtbsy10 at 7pt   \skewchar\sevensyb='60

\font\eightrm=mtr10 at 8pt
\font\eightbf=mtbx10 at 8pt
\font\eightit=mtti10 at 8pt
\font\eighti=mtmi10 at 8pt      \skewchar\eighti='177
\font\eightmib=mtmib10 at 8pt   \skewchar\eightmib='177
\font\eightsy=mtsy10 at 8pt     \skewchar\eightsy='60
\font\eightsyb=mtbsy10 at 8pt   \skewchar\eightsyb='60
\font\eightsl=mtsl10 at 8pt
\font\eighttt=mttt10 at 8pt     \hyphenchar\eighttt=-1
\font\eightcsc=mtcsc10 at 8pt
\font\eightsf=mtss10 at 8pt

\font\ninerm=mtr10 at 9pt
\font\ninebf=mtbx10 at 9pt
\font\nineit=mtti10 at 9pt
\font\ninei=mtmi10 at 9pt      \skewchar\ninei='177
\font\ninemib=mtmib10 at 9pt   \skewchar\ninemib='177
\font\ninesy=mtsy10 at 9pt     \skewchar\ninesy='60
\font\ninesyb=mtbsy10 at 9pt   \skewchar\ninesyb='60
\font\ninesl=mtsl10 at 9pt
\font\ninett=mttt10 at 9pt     \hyphenchar\ninett=-1
\font\ninecsc=mtcsc10 at 9pt
\font\ninesf=mtss10 at 9pt

\font\tenrm=mtr10
\font\tenbf=mtbx10
\font\tenit=mtti10
\font\teni=mtmi10		\skewchar\teni='177
\font\tenmib=mtmib10	\skewchar\tenmib='177
\font\tensy=mtsy10		\skewchar\tensy='60
\font\tensyb=mtbsy10	\skewchar\tensyb='60
\font\tenex=cmex10
\font\tensl=mtsl10
\font\tentt=mttt10		\hyphenchar\tentt=-1
\font\tencsc=mtcsc10
\font\tensf=mtss10

\font\elevenrm=mtr10 at 11pt
\font\elevenbf=mtbx10 at 11pt
\font\elevenit=mtti10 at 11pt
\font\eleveni=mtmi10 at 11pt      \skewchar\eleveni='177
\font\elevenmib=mtmib10 at 11pt   \skewchar\elevenmib='177
\font\elevensy=mtsy10 at 11pt     \skewchar\elevensy='60
\font\elevensyb=mtbsy10 at 11pt   \skewchar\elevensyb='60
\font\elevensl=mtsl10 at 11pt
\font\eleventt=mttt10 at 11pt     \hyphenchar\eleventt=-1
\font\elevencsc=mtcsc10 at 11pt
\font\elevensf=mtss10 at 11pt

\font\twelverm=mtr10 at 12pt
\font\twelvebf=mtbx10 at 12pt
\font\twelveit=mtti10 at 12pt
\font\twelvesl=mtsl10 at 12pt
\font\twelvett=mttt10 at 12pt     \hyphenchar\twelvett=-1
\font\twelvecsc=mtcsc10 at 12pt
\font\twelvesf=mtss10 at 12pt
\font\twelvei=mtmi10 at 12pt      \skewchar\twelvei='177
\font\twelvemib=mtmib10 at 12pt   \skewchar\twelvemib='177
\font\twelvesy=mtsy10 at 12pt     \skewchar\twelvesy='60
\font\twelvesyb=mtbsy10 at 12pt   \skewchar\twelvesyb='60

\font\fourteenrm=mtr10 at 14pt
\font\fourteenbf=mtbx10 at 14pt
\font\fourteenit=mtti10 at 14pt
\font\fourteeni=mtmi10 at 14pt      \skewchar\fourteeni='177
\font\fourteenmib=mtmib10 at 14pt   \skewchar\fourteenmib='177
\font\fourteensy=mtsy10 at 14pt     \skewchar\fourteensy='60
\font\fourteensyb=mtbsy10 at 14pt   \skewchar\fourteensyb='60
\font\fourteensl=mtsl10 at 14pt
\font\fourteentt=mttt10 at 14pt     \hyphenchar\fourteentt=-1
\font\fourteencsc=mtcsc10 at 14pt
\font\fourteensf=mtss10 at 14pt

\font\seventeenrm=mtr10 at 17pt
\font\seventeenbf=mtbx10 at 17pt
\font\seventeenit=mtti10 at 17pt
\font\seventeeni=mtmi10 at 17pt      \skewchar\seventeeni='177
\font\seventeenmib=mtmib10 at 17pt   \skewchar\seventeenmib='177
\font\seventeensy=mtsy10 at 17pt     \skewchar\seventeensy='60
\font\seventeensyb=mtbsy10 at 17pt   \skewchar\seventeensyb='60
\font\seventeensl=mtsl10 at 17pt
\font\seventeentt=mttt10 at 17pt     \hyphenchar\seventeentt=-1
\font\seventeencsc=mtcsc10 at 17pt
\font\seventeensf=mtss10 at 17pt


\newfam\xmfam
\newfam\ymfam

\font\fivexm=mtxm10 at 5pt
\font\sixxm=mtxm10 at 6pt
\font\sevenxm=mtxm10 at 7pt
\font\eightxm=mtxm10 at 8pt
\font\ninexm=mtxm10 at 9pt
\font\tenxm=mtxm10
\font\elevenxm=mtxm10 at 11pt
\font\twelvexm=mtxm10 at 12pt
\font\fourteenxm=mtxm10 at 14pt
\font\seventeenxm=mtxm10 at 17pt

\font\fiveym=mtym10 at 5pt
\font\sixym=mtym10 at 6pt
\font\sevenym=mtym10 at 7pt
\font\eightym=mtym10 at 8pt
\font\nineym=mtym10 at 9pt
\font\tenym=mtym10
\font\elevenym=mtym10 at 11pt
\font\twelveym=mtym10 at 12pt
\font\fourteenym=mtym10 at 14pt
\font\seventeenym=mtym10 at 17pt
\else
\font\fiverm=cmr5
\font\fivei=cmmi5             \skewchar\fivei='177
\font\fivemib=cmmib10 at 5pt  \skewchar\fivemib='177
\font\fivesy=cmsy5            \skewchar\fivesy='60
\font\fivesyb=cmbsy10 at 5pt  \skewchar\fivesyb='60
\font\fivebf=cmbx5

\font\sixrm=cmr6
\font\sixi=cmmi6             \skewchar\sixi='177
\font\sixmib=cmmib10 at 6pt  \skewchar\sixmib='177
\font\sixsy=cmsy6            \skewchar\sixsy='60
\font\sixsyb=cmbsy10 at 6pt  \skewchar\sixsyb='60
\font\sixbf=cmbx6

\font\sevenrm=cmr7
\font\seveni=cmmi7             \skewchar\seveni='177
\font\sevenmib=cmmib10 at 7pt  \skewchar\sevenmib='177
\font\sevensy=cmsy7            \skewchar\sevensy='60
\font\sevensyb=cmbsy10 at 7pt  \skewchar\sevensyb='60
\font\sevenbf=cmbx7

\font\eightrm=cmr8
\font\eightbf=cmbx8
\font\eightit=cmti8
\font\eighti=cmmi8			\skewchar\eighti='177
\font\eightmib=cmmib10 at 8pt	\skewchar\eightmib='177
\font\eightsy=cmsy8			\skewchar\eightsy='60
\font\eightsyb=cmbsy10 at 8pt	\skewchar\eightsyb='60
\font\eightsl=cmsl8
\font\eighttt=cmtt8			\hyphenchar\eighttt=-1
\font\eightcsc=cmcsc10 at 8pt
\font\eightsf=cmss8

\font\ninerm=cmr9
\font\ninebf=cmbx9
\font\nineit=cmti9
\font\ninei=cmmi9			\skewchar\ninei='177
\font\ninemib=cmmib10 at 9pt	\skewchar\ninemib='177
\font\ninesy=cmsy9			\skewchar\ninesy='60
\font\ninesyb=cmbsy10 at 9pt	\skewchar\ninesyb='60
\font\ninesl=cmsl9
\font\ninett=cmtt9			\hyphenchar\ninett=-1
\font\ninecsc=cmcsc10 at 9pt
\font\ninesf=cmss9

\font\tenrm=cmr10
\font\tenbf=cmbx10
\font\tenit=cmti10
\font\teni=cmmi10		\skewchar\teni='177
\font\tenmib=cmmib10	\skewchar\tenmib='177
\font\tensy=cmsy10		\skewchar\tensy='60
\font\tensyb=cmbsy10	\skewchar\tensyb='60
\font\tenex=cmex10
\font\tensl=cmsl10
\font\tentt=cmtt10		\hyphenchar\tentt=-1
\font\tencsc=cmcsc10
\font\tensf=cmss10

\font\elevenrm=cmr10 scaled \magstephalf
\font\elevenbf=cmbx10 scaled \magstephalf
\font\elevenit=cmti10 scaled \magstephalf
\font\eleveni=cmmi10 scaled \magstephalf	\skewchar\eleveni='177
\font\elevenmib=cmmib10 scaled \magstephalf	\skewchar\elevenmib='177
\font\elevensy=cmsy10 scaled \magstephalf	\skewchar\elevensy='60
\font\elevensyb=cmbsy10 scaled \magstephalf	\skewchar\elevensyb='60
\font\elevensl=cmsl10 scaled \magstephalf
\font\eleventt=cmtt10 scaled \magstephalf	\hyphenchar\eleventt=-1
\font\elevencsc=cmcsc10 scaled \magstephalf
\font\elevensf=cmss10 scaled \magstephalf

\font\twelverm=cmr10 scaled \magstep1
\font\twelvebf=cmbx10 scaled \magstep1
\font\twelvei=cmmi10 scaled \magstep1      \skewchar\twelvei='177
\font\twelvemib=cmmib10 scaled \magstep1   \skewchar\twelvemib='177
\font\twelvesy=cmsy10 scaled \magstep1     \skewchar\twelvesy='60
\font\twelvesyb=cmbsy10 scaled \magstep1   \skewchar\twelvesyb='60

\font\fourteenrm=cmr10 scaled \magstep2
\font\fourteenbf=cmbx10 scaled \magstep2
\font\fourteenit=cmti10 scaled \magstep2
\font\fourteeni=cmmi10 scaled \magstep2		\skewchar\fourteeni='177
\font\fourteenmib=cmmib10 scaled \magstep2	\skewchar\fourteenmib='177
\font\fourteensy=cmsy10 scaled \magstep2	\skewchar\fourteensy='60
\font\fourteensyb=cmbsy10 scaled \magstep2	\skewchar\fourteensyb='60
\font\fourteensl=cmsl10 scaled \magstep2
\font\fourteentt=cmtt10 scaled \magstep2	\hyphenchar\fourteentt=-1
\font\fourteencsc=cmcsc10 scaled \magstep2
\font\fourteensf=cmss10 scaled \magstep2

\font\seventeenrm=cmr10 scaled \magstep3
\font\seventeenbf=cmbx10 scaled \magstep3
\font\seventeenit=cmti10 scaled \magstep3
\font\seventeeni=cmmi10 scaled \magstep3	\skewchar\seventeeni='177
\font\seventeenmib=cmmib10 scaled \magstep3	\skewchar\seventeenmib='177
\font\seventeensy=cmsy10 scaled \magstep3	\skewchar\seventeensy='60
\font\seventeensyb=cmbsy10 scaled \magstep3	\skewchar\seventeensyb='60
\font\seventeensl=cmsl10 scaled \magstep3
\font\seventeentt=cmtt10 scaled \magstep3	\hyphenchar\seventeentt=-1
\font\seventeencsc=cmcsc10 scaled \magstep3
\font\seventeensf=cmss10 scaled \magstep3
\fi

\def\hexnumber#1{\ifcase#1 0\or1\or2\or3\or4\or5\or6\or7\or8\or9\or
  A\or B\or C\or D\or E\or F\fi}

\ifprod@font
  \edef\@xm{\hexnumber\xmfam}
  \edef\@ym{\hexnumber\ymfam}
\fi

\def\makestrut{%
  \setbox\strutbox=\hbox{%
    \vrule height.7\baselineskip depth.3\baselineskip width \z@}%
}

\def\baselinestretch{1}
\newskip\tmp@bls

\def\b@ls#1{
  \tmp@bls=#1\relax
  \baselineskip=#1\relax\makestrut
  \normalbaselineskip=\baselinestretch\tmp@bls
  \normalbaselines
}

\def\nostb@ls#1{
  \normalbaselineskip=#1\relax
  \normalbaselines
  \makestrut
}

%

\newfam\mibfam 
\newfam\sybfam 
\newfam\scfam  
\newfam\sffam  

\def\mit{\fam\@ne}

\def\cal{\fam\tw@}

\def\em{\ifdim\fontdimen1\font>\z@ \rm\else\it\fi}

\textfont3=\tenex
\scriptfont3=\tenex
\scriptscriptfont3=\tenex

\setbox0=\hbox{\tenex B} \p@renwd=\wd0 

\def\eightpoint{
  \def\rm{\fam0\eightrm}%
  \textfont0=\eightrm \scriptfont0=\sixrm \scriptscriptfont0=\fiverm%
  \textfont1=\eighti  \scriptfont1=\sixi  \scriptscriptfont1=\fivei%
  \textfont2=\eightsy \scriptfont2=\sixsy \scriptscriptfont2=\fivesy%
  \textfont\itfam=\eightit\def\it{\fam\itfam\eightit}%
  \ifprod@font
    \scriptfont\itfam=\sixit
      \scriptscriptfont\itfam=\fiveit
  \else
    \scriptfont\itfam=\eightit
      \scriptscriptfont\itfam=\eightit
  \fi
  \textfont\bffam=\eightbf%
    \scriptfont\bffam=\sixbf%
      \scriptscriptfont\bffam=\fivebf%
  \def\bf{\fam\bffam\eightbf}%
  \textfont\slfam=\eightsl\def\sl{\fam\slfam\eightsl}%
  \ifprod@font
    \scriptfont\slfam=\sixsl
      \scriptscriptfont\slfam=\fivesl
  \else
    \scriptfont\slfam=\eightsl
      \scriptscriptfont\slfam=\eightsl
  \fi
  \textfont\ttfam=\eighttt\def\tt{\fam\ttfam\eighttt}%
  \ifprod@font
    \scriptfont\ttfam=\sixtt
      \scriptscriptfont\ttfam=\fivett
  \else
    \scriptfont\ttfam=\eighttt
      \scriptscriptfont\ttfam=\eighttt
  \fi
  \textfont\scfam=\eightcsc\def\sc{\fam\scfam\eightcsc}%
  \ifprod@font
    \scriptfont\scfam=\sixcsc
      \scriptscriptfont\scfam=\fivecsc
  \else
    \scriptfont\scfam=\eightcsc
      \scriptscriptfont\scfam=\eightcsc
  \fi
  \textfont\sffam=\eightsf\def\sf{\fam\sffam\eightsf}%
  \ifprod@font
    \scriptfont\sffam=\sixsf
      \scriptscriptfont\sffam=\fivesf
  \else
    \scriptfont\sffam=\eightsf
      \scriptscriptfont\sffam=\eightsf
  \fi
  \textfont\mibfam=\eightmib
    \scriptfont\mibfam=\sixmib
      \scriptscriptfont\mibfam=\fivemib
  \textfont\sybfam=\eightsyb
    \scriptfont\sybfam=\sixsyb
      \scriptscriptfont\sybfam=\fivesyb
  \ifprod@font
    \textfont\xmfam=\eightxm
      \scriptfont\xmfam=\sixxm
        \scriptscriptfont\xmfam=\fivexm
    \textfont\ymfam=\eightym
      \scriptfont\ymfam=\sixym
        \scriptscriptfont\ymfam=\fiveym
  \fi
  \def\oldstyle{\fam\@ne\eighti}%
  \def\boldstyle{\fam\mibfam\eightmib}%
  \b@ls{10pt}\rm%
}

\def\ninepoint{
  \def\rm{\fam0\ninerm}%
  \textfont0=\ninerm \scriptfont0=\sixrm \scriptscriptfont0=\fiverm%
  \textfont1=\ninei  \scriptfont1=\sixi  \scriptscriptfont1=\fivei%
  \textfont2=\ninesy \scriptfont2=\sixsy \scriptscriptfont2=\fivesy%
  \textfont\itfam=\nineit\def\it{\fam\itfam\nineit}%
  \ifprod@font
    \scriptfont\itfam=\sixit
      \scriptscriptfont\itfam=\fiveit
  \else
    \scriptfont\itfam=\nineit
      \scriptscriptfont\itfam=\nineit
  \fi
  \textfont\bffam=\ninebf%
    \scriptfont\bffam=\sixbf%
      \scriptscriptfont\bffam=\fivebf%
  \def\bf{\fam\bffam\ninebf}%
  \textfont\slfam=\ninesl\def\sl{\fam\slfam\ninesl}%
  \ifprod@font
    \scriptfont\slfam=\sixsl
      \scriptscriptfont\slfam=\fivesl
  \else
    \scriptfont\slfam=\ninesl
      \scriptscriptfont\slfam=\ninesl
  \fi
  \textfont\ttfam=\ninett\def\tt{\fam\ttfam\ninett}%
  \ifprod@font
    \scriptfont\ttfam=\sixtt
      \scriptscriptfont\ttfam=\fivett
  \else
    \scriptfont\ttfam=\ninett
      \scriptscriptfont\ttfam=\ninett
  \fi
  \textfont\scfam=\ninecsc\def\sc{\fam\scfam\ninecsc}%
  \ifprod@font
    \scriptfont\scfam=\sixcsc
      \scriptscriptfont\scfam=\fivecsc
  \else
    \scriptfont\scfam=\ninecsc
      \scriptscriptfont\scfam=\ninecsc
  \fi
  \textfont\sffam=\ninesf\def\sf{\fam\sffam\ninesf}%
  \ifprod@font
    \scriptfont\sffam=\sixsf
      \scriptscriptfont\sffam=\fivesf
  \else
    \scriptfont\sffam=\ninesf
      \scriptscriptfont\sffam=\ninesf
  \fi
  \textfont\mibfam=\ninemib
    \scriptfont\mibfam=\sixmib
      \scriptscriptfont\mibfam=\fivemib
  \textfont\sybfam=\ninesyb
    \scriptfont\sybfam=\sixsyb
      \scriptscriptfont\sybfam=\fivesyb
  \ifprod@font
    \textfont\xmfam=\ninexm
      \scriptfont\xmfam=\sixxm
        \scriptscriptfont\xmfam=\fivexm
    \textfont\ymfam=\nineym
      \scriptfont\ymfam=\sixym
        \scriptscriptfont\ymfam=\fiveym
  \fi
  \def\oldstyle{\fam\@ne\ninei}%
  \def\boldstyle{\fam\mibfam\ninemib}%
  \b@ls{\TextLeading plus \Feathering}\rm%
}

\def\tenpoint{
  \def\rm{\fam0\tenrm}%
  \textfont0=\tenrm \scriptfont0=\sevenrm \scriptscriptfont0=\fiverm%
  \textfont1=\teni  \scriptfont1=\seveni  \scriptscriptfont1=\fivei%
  \textfont2=\tensy \scriptfont2=\sevensy \scriptscriptfont2=\fivesy%
  \textfont\itfam=\tenit\def\it{\fam\itfam\tenit}%
  \ifprod@font
    \scriptfont\itfam=\sevenit
      \scriptscriptfont\itfam=\fiveit
  \else
    \scriptfont\itfam=\tenit
      \scriptscriptfont\itfam=\tenit
  \fi
  \textfont\bffam=\tenbf%
    \scriptfont\bffam=\sevenbf%
      \scriptscriptfont\bffam=\fivebf%
  \def\bf{\fam\bffam\tenbf}%
  \textfont\slfam=\tensl\def\sl{\fam\slfam\tensl}%
  \ifprod@font
    \scriptfont\slfam=\sevensl
      \scriptscriptfont\slfam=\fivesl
  \else
    \scriptfont\slfam=\tensl
      \scriptscriptfont\slfam=\tensl
  \fi
  \textfont\ttfam=\tentt\def\tt{\fam\ttfam\tentt}%
  \ifprod@font
    \scriptfont\ttfam=\seventt
      \scriptscriptfont\ttfam=\fivett
  \else
    \scriptfont\ttfam=\tentt
      \scriptscriptfont\ttfam=\tentt
  \fi
  \textfont\scfam=\tencsc\def\sc{\fam\scfam\tencsc}%
  \ifprod@font
    \scriptfont\scfam=\sevencsc
      \scriptscriptfont\scfam=\fivecsc
  \else
    \scriptfont\scfam=\tencsc
      \scriptscriptfont\scfam=\tencsc
  \fi
  \textfont\sffam=\tensf\def\sf{\fam\sffam\tensf}%
  \ifprod@font
    \scriptfont\sffam=\sevensf
      \scriptscriptfont\sffam=\fivesf
  \else
    \scriptfont\sffam=\tensf
      \scriptscriptfont\sffam=\tensf
  \fi
  \textfont\mibfam=\tenmib
    \scriptfont\mibfam=\sevenmib
      \scriptscriptfont\mibfam=\fivemib
  \textfont\sybfam=\tensyb
    \scriptfont\sybfam=\sevensyb
      \scriptscriptfont\sybfam=\fivesyb
  \ifprod@font
    \textfont\xmfam=\tenxm
      \scriptfont\xmfam=\sevenxm
        \scriptscriptfont\xmfam=\fivexm
    \textfont\ymfam=\tenym
      \scriptfont\ymfam=\sevenym
        \scriptscriptfont\ymfam=\fiveym
  \fi
  \def\oldstyle{\fam\@ne\teni}%
  \def\boldstyle{\fam\mibfam\tenmib}%
  \b@ls{11pt}\rm%
}

\def\elevenpoint{
  \def\rm{\fam0\elevenrm}%
  \textfont0=\elevenrm \scriptfont0=\eightrm \scriptscriptfont0=\sixrm%
  \textfont1=\eleveni  \scriptfont1=\eighti  \scriptscriptfont1=\sixi%
  \textfont2=\elevensy \scriptfont2=\eightsy \scriptscriptfont2=\sixsy%
  \textfont\itfam=\elevenit\def\it{\fam\itfam\elevenit}%
  \ifprod@font
    \scriptfont\itfam=\eightit
      \scriptscriptfont\itfam=\sixit
  \else
    \scriptfont\itfam=\elevenit
      \scriptscriptfont\itfam=\elevenit
  \fi
  \textfont\bffam=\elevenbf%
    \scriptfont\bffam=\eightbf%
      \scriptscriptfont\bffam=\sixbf%
  \def\bf{\fam\bffam\elevenbf}%
  \textfont\slfam=\elevensl\def\sl{\fam\slfam\elevensl}%
  \ifprod@font
    \scriptfont\slfam=\eightsl
      \scriptscriptfont\slfam=\sixsl
  \else
    \scriptfont\slfam=\elevensl
      \scriptscriptfont\slfam=\elevensl
  \fi
  \textfont\ttfam=\eleventt\def\tt{\fam\ttfam\eleventt}%
  \ifprod@font
    \scriptfont\ttfam=\eighttt
      \scriptscriptfont\ttfam=\sixtt
  \else
    \scriptfont\ttfam=\eleventt
      \scriptscriptfont\ttfam=\eleventt
  \fi
  \textfont\scfam=\elevencsc\def\sc{\fam\scfam\elevencsc}%
  \ifprod@font
    \scriptfont\scfam=\eightcsc
      \scriptscriptfont\scfam=\sixcsc
  \else
    \scriptfont\scfam=\elevencsc
      \scriptscriptfont\scfam=\elevencsc
  \fi
  \textfont\sffam=\elevensf\def\sf{\fam\sffam\elevensf}%
  \ifprod@font
    \scriptfont\sffam=\eightsf
      \scriptscriptfont\sffam=\sixsf
  \else
    \scriptfont\sffam=\elevensf
      \scriptscriptfont\sffam=\elevensf
  \fi
  \textfont\mibfam=\elevenmib
    \scriptfont\mibfam=\eightmib
      \scriptscriptfont\mibfam=\sixmib
  \textfont\sybfam=\elevensyb
    \scriptfont\sybfam=\eightsyb
      \scriptscriptfont\sybfam=\sixsyb
  \ifprod@font
    \textfont\xmfam=\elevenxm
      \scriptfont\xmfam=\eightxm
       \scriptscriptfont\xmfam=\sixxm
    \textfont\ymfam=\elevenym
      \scriptfont\ymfam=\eightym
        \scriptscriptfont\ymfam=\sixym
   \fi
  \def\oldstyle{\fam\@ne\eleveni}%
  \def\boldstyle{\fam\mibfam\elevenmib}%
  \b@ls{13pt}\rm%
}

\def\fourteenpoint{
  \def\rm{\fam0\fourteenrm}%
  \textfont0\fourteenrm  \scriptfont0\tenrm  \scriptscriptfont0\sevenrm%
  \textfont1\fourteeni   \scriptfont1\teni   \scriptscriptfont1\seveni%
  \textfont2\fourteensy  \scriptfont2\tensy  \scriptscriptfont2\sevensy%
  \textfont\itfam=\fourteenit\def\it{\fam\itfam\fourteenit}%
  \ifprod@font
    \scriptfont\itfam=\tenit
      \scriptscriptfont\itfam=\sevenit
  \else
    \scriptfont\itfam=\fourteenit
      \scriptscriptfont\itfam=\fourteenit
  \fi
  \textfont\bffam=\fourteenbf%
    \scriptfont\bffam=\tenbf%
      \scriptscriptfont\bffam=\sevenbf%
  \def\bf{\fam\bffam\fourteenbf}%
  \textfont\slfam=\fourteensl\def\sl{\fam\slfam\fourteensl}%
  \ifprod@font
    \scriptfont\slfam=\tensl
      \scriptscriptfont\slfam=\sevensl
  \else
    \scriptfont\slfam=\fourteensl
      \scriptscriptfont\slfam=\fourteensl
  \fi
  \textfont\ttfam=\fourteentt\def\tt{\fam\ttfam\fourteentt}%
  \ifprod@font
    \scriptfont\ttfam=\tentt
      \scriptscriptfont\ttfam=\seventt
  \else
    \scriptfont\ttfam=\fourteentt
      \scriptscriptfont\ttfam=\fourteentt
  \fi
  \textfont\scfam=\fourteencsc\def\sc{\fam\scfam\fourteencsc}%
  \ifprod@font
    \scriptfont\scfam=\tencsc
      \scriptscriptfont\scfam=\sevencsc
  \else
    \scriptfont\scfam=\fourteencsc
      \scriptscriptfont\scfam=\fourteencsc
  \fi
  \textfont\sffam=\fourteensf\def\sf{\fam\sffam\fourteensf}%
  \ifprod@font
    \scriptfont\sffam=\tensf
      \scriptscriptfont\sffam=\sevensf
  \else
    \scriptfont\sffam=\fourteensf
      \scriptscriptfont\sffam=\fourteensf
  \fi
  \textfont\mibfam=\fourteenmib
    \scriptfont\mibfam=\tenmib
      \scriptscriptfont\mibfam=\sevenmib
  \textfont\sybfam=\fourteensyb
    \scriptfont\sybfam=\tensyb
      \scriptscriptfont\sybfam=\sevensyb
  \ifprod@font
    \textfont\xmfam=\fourteenxm
      \scriptfont\xmfam=\tenxm
        \scriptscriptfont\xmfam=\sevenxm
   \textfont\ymfam=\fourteenym
      \scriptfont\ymfam=\tenym
        \scriptscriptfont\ymfam=\sevenym
  \fi
  \def\oldstyle{\fam\@ne\fourteeni}%
  \def\boldstyle{\fam\mibfam\fourteenmib}%
  \b@ls{17pt}\rm%
}

\def\seventeenpoint{
  \def\rm{\fam0\seventeenrm}%
  \textfont0\seventeenrm  \scriptfont0\twelverm  \scriptscriptfont0\tenrm%
  \textfont1\seventeeni   \scriptfont1\twelvei   \scriptscriptfont1\teni%
  \textfont2\seventeensy  \scriptfont2\twelvesy  \scriptscriptfont2\tensy%
  \textfont\itfam=\seventeenit\def\it{\fam\itfam\seventeenit}%
  \ifprod@font
    \scriptfont\itfam=\twelveit
      \scriptscriptfont\itfam=\tenit
  \else
    \scriptfont\itfam=\seventeenit
      \scriptscriptfont\itfam=\seventeenit
  \fi
  \textfont\bffam=\seventeenbf%
    \scriptfont\bffam=\twelvebf%
      \scriptscriptfont\bffam=\tenbf%
  \def\bf{\fam\bffam\seventeenbf}%
  \textfont\slfam=\seventeensl\def\sl{\fam\slfam\seventeensl}%
  \ifprod@font
    \scriptfont\slfam=\twelvesl
      \scriptscriptfont\slfam=\tensl
  \else
    \scriptfont\slfam=\seventeensl
      \scriptscriptfont\slfam=\seventeensl
  \fi
  \textfont\ttfam=\seventeentt\def\tt{\fam\ttfam\seventeentt}%
  \ifprod@font
    \scriptfont\ttfam=\twelvett
      \scriptscriptfont\ttfam=\tentt
  \else
    \scriptfont\ttfam=\seventeentt
      \scriptscriptfont\ttfam=\seventeentt
  \fi
  \textfont\scfam=\seventeencsc\def\sc{\fam\scfam\seventeencsc}%
  \ifprod@font
    \scriptfont\scfam=\twelvecsc
      \scriptscriptfont\scfam=\tencsc
  \else
    \scriptfont\scfam=\seventeencsc
      \scriptscriptfont\scfam=\seventeencsc
  \fi
  \textfont\sffam=\seventeensf\def\sf{\fam\sffam\seventeensf}%
  \ifprod@font
    \scriptfont\sffam=\twelvesf
      \scriptscriptfont\sffam=\tensf
  \else
    \scriptfont\sffam=\seventeensf
      \scriptscriptfont\sffam=\seventeensf
  \fi
  \textfont\mibfam=\seventeenmib
    \scriptfont\mibfam=\twelvemib
      \scriptscriptfont\mibfam=\tenmib
  \textfont\sybfam=\seventeensyb
    \scriptfont\sybfam=\twelvesyb
      \scriptscriptfont\sybfam=\tensyb
  \ifprod@font
    \textfont\xmfam=\seventeenxm
      \scriptfont\xmfam=\twelvexm
        \scriptscriptfont\xmfam=\tenxm
    \textfont\ymfam=\seventeenym
      \scriptfont\ymfam=\twelveym
        \scriptscriptfont\ymfam=\tenym
  \fi
  \def\oldstyle{\fam\@ne\seventeeni}%
  \def\boldstyle{\fam\mibfam\seventeenmib}%
  \b@ls{20pt}\rm%
}

\lineskip=1pt      \normallineskip=\lineskip
\lineskiplimit=\z@ \normallineskiplimit=\lineskiplimit



\def\la{\mathrel{\mathchoice {\vcenter{\offinterlineskip\halign{\hfil
$\displaystyle##$\hfil\cr<\cr\sim\cr}}}
{\vcenter{\offinterlineskip\halign{\hfil$\textstyle##$\hfil\cr
<\cr\sim\cr}}}
{\vcenter{\offinterlineskip\halign{\hfil$\scriptstyle##$\hfil\cr
<\cr\sim\cr}}}
{\vcenter{\offinterlineskip\halign{\hfil$\scriptscriptstyle##$\hfil\cr
<\cr\sim\cr}}}}}

\def\ga{\mathrel{\mathchoice {\vcenter{\offinterlineskip\halign{\hfil
$\displaystyle##$\hfil\cr>\cr\sim\cr}}}
{\vcenter{\offinterlineskip\halign{\hfil$\textstyle##$\hfil\cr
>\cr\sim\cr}}}
{\vcenter{\offinterlineskip\halign{\hfil$\scriptstyle##$\hfil\cr
>\cr\sim\cr}}}
{\vcenter{\offinterlineskip\halign{\hfil$\scriptscriptstyle##$\hfil\cr
>\cr\sim\cr}}}}}

\def\getsto{\mathrel{\mathchoice {\vcenter{\offinterlineskip
\halign{\hfil
$\displaystyle##$\hfil\cr\gets\cr\to\cr}}}
{\vcenter{\offinterlineskip\halign{\hfil$\textstyle##$\hfil\cr\gets
\cr\to\cr}}}
{\vcenter{\offinterlineskip\halign{\hfil$\scriptstyle##$\hfil\cr\gets
\cr\to\cr}}}
{\vcenter{\offinterlineskip\halign{\hfil$\scriptscriptstyle##$\hfil\cr
\gets\cr\to\cr}}}}}

\def\lid{\mathrel{\mathchoice {\vcenter{\offinterlineskip\halign{\hfil
$\displaystyle##$\hfil\cr<\cr\noalign{\vskip1.2pt}=\cr}}}
{\vcenter{\offinterlineskip\halign{\hfil$\textstyle##$\hfil\cr<\cr
\noalign{\vskip1.2pt}=\cr}}}
{\vcenter{\offinterlineskip\halign{\hfil$\scriptstyle##$\hfil\cr<\cr
\noalign{\vskip1pt}=\cr}}}
{\vcenter{\offinterlineskip\halign{\hfil$\scriptscriptstyle##$\hfil\cr
<\cr
\noalign{\vskip0.9pt}=\cr}}}}}

\def\gid{\mathrel{\mathchoice {\vcenter{\offinterlineskip\halign{\hfil
$\displaystyle##$\hfil\cr>\cr\noalign{\vskip1.2pt}=\cr}}}
{\vcenter{\offinterlineskip\halign{\hfil$\textstyle##$\hfil\cr>\cr
\noalign{\vskip1.2pt}=\cr}}}
{\vcenter{\offinterlineskip\halign{\hfil$\scriptstyle##$\hfil\cr>\cr
\noalign{\vskip1pt}=\cr}}}
{\vcenter{\offinterlineskip\halign{\hfil$\scriptscriptstyle##$\hfil\cr
>\cr
\noalign{\vskip0.9pt}=\cr}}}}}

\def\grole{\mathrel{\mathchoice {\vcenter{\offinterlineskip\halign{\hfil
$\displaystyle##$\hfil\cr>\cr\noalign{\vskip-1.5pt}<\cr}}}
{\vcenter{\offinterlineskip\halign{\hfil$\textstyle##$\hfil\cr
>\cr\noalign{\vskip-1.5pt}<\cr}}}
{\vcenter{\offinterlineskip\halign{\hfil$\scriptstyle##$\hfil\cr
>\cr\noalign{\vskip-1pt}<\cr}}}
{\vcenter{\offinterlineskip\halign{\hfil$\scriptscriptstyle##$\hfil\cr
>\cr\noalign{\vskip-0.5pt}<\cr}}}}}

\def\leogr{\mathrel{\mathchoice {\vcenter{\offinterlineskip\halign{\hfil
$\displaystyle##$\hfil\cr<\cr\noalign{\vskip-1.5pt}>\cr}}}
{\vcenter{\offinterlineskip\halign{\hfil$\textstyle##$\hfil\cr
<\cr\noalign{\vskip-1.5pt}>\cr}}}
{\vcenter{\offinterlineskip\halign{\hfil$\scriptstyle##$\hfil\cr
<\cr\noalign{\vskip-1pt}>\cr}}}
{\vcenter{\offinterlineskip\halign{\hfil$\scriptscriptstyle##$\hfil\cr
<\cr\noalign{\vskip-0.5pt}>\cr}}}}}

\def\loa{\mathrel{\mathchoice {\vcenter{\offinterlineskip\halign{\hfil
$\displaystyle##$\hfil\cr<\cr\approx\cr}}}
{\vcenter{\offinterlineskip\halign{\hfil$\textstyle##$\hfil\cr
<\cr\approx\cr}}}
{\vcenter{\offinterlineskip\halign{\hfil$\scriptstyle##$\hfil\cr
<\cr\approx\cr}}}
{\vcenter{\offinterlineskip\halign{\hfil$\scriptscriptstyle##$\hfil\cr
<\cr\approx\cr}}}}}

\def\goa{\mathrel{\mathchoice {\vcenter{\offinterlineskip\halign{\hfil
$\displaystyle##$\hfil\cr>\cr\approx\cr}}}
{\vcenter{\offinterlineskip\halign{\hfil$\textstyle##$\hfil\cr
>\cr\approx\cr}}}
{\vcenter{\offinterlineskip\halign{\hfil$\scriptstyle##$\hfil\cr
>\cr\approx\cr}}}
{\vcenter{\offinterlineskip\halign{\hfil$\scriptscriptstyle##$\hfil\cr
>\cr\approx\cr}}}}}

\def\diameter{{\ifmmode\mathchoice
{\ooalign{\hfil\hbox{$\displaystyle/$}\hfil\crcr
{\hbox{$\displaystyle\mathchar"20D$}}}}
{\ooalign{\hfil\hbox{$\textstyle/$}\hfil\crcr
{\hbox{$\textstyle\mathchar"20D$}}}}
{\ooalign{\hfil\hbox{$\scriptstyle/$}\hfil\crcr
{\hbox{$\scriptstyle\mathchar"20D$}}}}
{\ooalign{\hfil\hbox{$\scriptscriptstyle/$}\hfil\crcr
{\hbox{$\scriptscriptstyle\mathchar"20D$}}}}
\else{\ooalign{\hfil/\hfil\crcr\mathhexbox20D}}%
\fi}}

\def\sq{\ifmmode\squareforqed\else{\unskip\nobreak\hfil
\penalty50\hskip1em\null\nobreak\hfil\squareforqed
\parfillskip=0pt\finalhyphendemerits=0\endgraf}\fi}
\def\squareforqed{\hbox{\rlap{$\sqcap$}$\sqcup$}}


\def\bbbc{{\mathchoice {\setbox0=\hbox{$\displaystyle\rm C$}\hbox{\hbox
to0pt{\kern0.4\wd0\vrule height0.9\ht0\hss}\box0}}
{\setbox0=\hbox{$\textstyle\rm C$}\hbox{\hbox
to0pt{\kern0.4\wd0\vrule height0.9\ht0\hss}\box0}}
{\setbox0=\hbox{$\scriptstyle\rm C$}\hbox{\hbox
to0pt{\kern0.4\wd0\vrule height0.9\ht0\hss}\box0}}
{\setbox0=\hbox{$\scriptscriptstyle\rm C$}\hbox{\hbox
to0pt{\kern0.4\wd0\vrule height0.9\ht0\hss}\box0}}}}
\def\bbbq{{\mathchoice {\setbox0=\hbox{$\displaystyle\rm
Q$}\hbox{\raise
0.15\ht0\hbox to0pt{\kern0.4\wd0\vrule height0.8\ht0\hss}\box0}}
{\setbox0=\hbox{$\textstyle\rm Q$}\hbox{\raise
0.15\ht0\hbox to0pt{\kern0.4\wd0\vrule height0.8\ht0\hss}\box0}}
{\setbox0=\hbox{$\scriptstyle\rm Q$}\hbox{\raise
0.15\ht0\hbox to0pt{\kern0.4\wd0\vrule height0.7\ht0\hss}\box0}}
{\setbox0=\hbox{$\scriptscriptstyle\rm Q$}\hbox{\raise
0.15\ht0\hbox to0pt{\kern0.4\wd0\vrule height0.7\ht0\hss}\box0}}}}
\def\bbbt{{\mathchoice {\setbox0=\hbox{$\displaystyle\rm
T$}\hbox{\hbox to0pt{\kern0.3\wd0\vrule height0.9\ht0\hss}\box0}}
{\setbox0=\hbox{$\textstyle\rm T$}\hbox{\hbox
to0pt{\kern0.3\wd0\vrule height0.9\ht0\hss}\box0}}
{\setbox0=\hbox{$\scriptstyle\rm T$}\hbox{\hbox
to0pt{\kern0.3\wd0\vrule height0.9\ht0\hss}\box0}}
{\setbox0=\hbox{$\scriptscriptstyle\rm T$}\hbox{\hbox
to0pt{\kern0.3\wd0\vrule height0.9\ht0\hss}\box0}}}}
\def\bbbs{{\mathchoice
{\setbox0=\hbox{$\displaystyle     \rm S$}\hbox{\raise0.5\ht0\hbox
to0pt{\kern0.35\wd0\vrule height0.45\ht0\hss}\hbox
to0pt{\kern0.55\wd0\vrule height0.5\ht0\hss}\box0}}
{\setbox0=\hbox{$\textstyle        \rm S$}\hbox{\raise0.5\ht0\hbox
to0pt{\kern0.35\wd0\vrule height0.45\ht0\hss}\hbox
to0pt{\kern0.55\wd0\vrule height0.5\ht0\hss}\box0}}
{\setbox0=\hbox{$\scriptstyle      \rm S$}\hbox{\raise0.5\ht0\hbox
to0pt{\kern0.35\wd0\vrule height0.45\ht0\hss}\raise0.05\ht0\hbox
to0pt{\kern0.5\wd0\vrule height0.45\ht0\hss}\box0}}
{\setbox0=\hbox{$\scriptscriptstyle\rm S$}\hbox{\raise0.5\ht0\hbox
to0pt{\kern0.4\wd0\vrule height0.45\ht0\hss}\raise0.05\ht0\hbox
to0pt{\kern0.55\wd0\vrule height0.45\ht0\hss}\box0}}}}
\def\bbbz{{\mathchoice {\hbox{$\sf\textstyle Z\kern-0.4em Z$}}
{\hbox{$\sf\textstyle Z\kern-0.4em Z$}}
{\hbox{$\sf\scriptstyle Z\kern-0.3em Z$}}
{\hbox{$\sf\scriptscriptstyle Z\kern-0.2em Z$}}}}


\ifprod@font
  \mathchardef\la="3\@xm2E
  \mathchardef\getsto="3\@xm1C
  \mathchardef\lid="3\@xm35
  \mathchardef\grole="3\@xm3F
  \mathchardef\loa="3\@xm2F
  \mathchardef\ga="3\@xm26
  \mathchardef\gid="3\@xm3D
  \mathchardef\leogr="3\@xm37
  \mathchardef\goa="3\@xm27
  \mathchardef\sq="0\@xm03
%
%
\def\diameter{{%
  \ifmmode
    \mathchoice
    {\ooalign{\hfil\hbox{$\displaystyle/$}\hfil\crcr
    {\lower.2ex\hbox{$\displaystyle\mathchar"20D$}}}}%
    {\ooalign{\hfil\hbox{$\textstyle/$}\hfil\crcr
    {\lower.2ex\hbox{$\textstyle\mathchar"20D$}}}}%
    {\ooalign{\hfil\hbox{$\scriptstyle/$}\hfil\crcr
    {\lower.1ex\hbox{$\scriptstyle\mathchar"20D$}}}}%
    {\ooalign{\hfil\hbox{$\scriptscriptstyle/$}\hfil\crcr
    {\lower.1ex\hbox{$\scriptscriptstyle\mathchar"20D$}}}}%
  \else
    {\ooalign{\hfil/\hfil\crcr\lower.2ex\hbox{\mathhexbox20D}}}%
  \fi
}}
%
%

\def\bbbc{{\Bbb{C}}}
\def\bbbq{{\Bbb{Q}}}
\def\bbbt{{\Bbb{T}}}
\def\bbbs{{\Bbb{S}}}
\def\bbbz{{\Bbb{Z}}}
\fi


\ifprod@font
\mathchardef\boxdot="2\@xm00
\mathchardef\boxplus="2\@xm01
\mathchardef\boxtimes="2\@xm02
\mathchardef\square="0\@xm03
\mathchardef\blacksquare="0\@xm04
\mathchardef\centerdot="2\@xm05
\mathchardef\lozenge="0\@xm06
\mathchardef\blacklozenge="0\@xm07
\mathchardef\circlearrowright="3\@xm08
\mathchardef\circlearrowleft="3\@xm09
\mathchardef\rightleftharpoons="3\@xm0A
\mathchardef\leftrightharpoons="3\@xm0B
\mathchardef\boxminus="2\@xm0C
\mathchardef\Vdash="3\@xm0D
\mathchardef\Vvdash="3\@xm0E
\mathchardef\vDash="3\@xm0F
\mathchardef\twoheadrightarrow="3\@xm10
\mathchardef\twoheadleftarrow="3\@xm11
\mathchardef\leftleftarrows="3\@xm12
\mathchardef\rightrightarrows="3\@xm13
\mathchardef\upuparrows="3\@xm14
\mathchardef\downdownarrows="3\@xm15
\mathchardef\upharpoonright="3\@xm16

\mathchardef\downharpoonright="3\@xm17
\mathchardef\upharpoonleft="3\@xm18
\mathchardef\downharpoonleft="3\@xm19
\mathchardef\rightarrowtail="3\@xm1A
\mathchardef\leftarrowtail="3\@xm1B
\mathchardef\leftrightarrows="3\@xm1C
\mathchardef\rightleftarrows="3\@xm1D
\mathchardef\Lsh="3\@xm1E
\mathchardef\Rsh="3\@xm1F
\mathchardef\rightsquigarrow="3\@xm20
\mathchardef\leftrightsquigarrow="3\@xm21
\mathchardef\looparrowleft="3\@xm22
\mathchardef\looparrowright="3\@xm23
\mathchardef\circeq="3\@xm24
\mathchardef\succsim="3\@xm25
\mathchardef\gtrsim="3\@xm26
\mathchardef\gtrapprox="3\@xm27
\mathchardef\multimap="3\@xm28
\mathchardef\therefore="3\@xm29
\mathchardef\because="3\@xm2A
\mathchardef\doteqdot="3\@xm2B

\mathchardef\triangleq="3\@xm2C
\mathchardef\precsim="3\@xm2D
\mathchardef\lesssim="3\@xm2E
\mathchardef\lessapprox="3\@xm2F
\mathchardef\eqslantless="3\@xm30
\mathchardef\eqslantgtr="3\@xm31
\mathchardef\curlyeqprec="3\@xm32
\mathchardef\curlyeqsucc="3\@xm33
\mathchardef\preccurlyeq="3\@xm34
\mathchardef\leqq="3\@xm35
\mathchardef\leqslant="3\@xm36
\mathchardef\lessgtr="3\@xm37
\mathchardef\backprime="0\@xm38
\mathchardef\risingdotseq="3\@xm3A
\mathchardef\fallingdotseq="3\@xm3B
\mathchardef\succcurlyeq="3\@xm3C
\mathchardef\geqq="3\@xm3D
\mathchardef\geqslant="3\@xm3E
\mathchardef\gtrless="3\@xm3F
\mathchardef\sqsubset="3\@xm40
\mathchardef\sqsupset="3\@xm41
\mathchardef\vartriangleright="3\@xm42
\mathchardef\vartriangleleft="3\@xm43
\mathchardef\trianglerighteq="3\@xm44
\mathchardef\trianglelefteq="3\@xm45
\mathchardef\bigstar="0\@xm46
\mathchardef\between="3\@xm47
\mathchardef\blacktriangledown="0\@xm48
\mathchardef\blacktriangleright="3\@xm49
\mathchardef\blacktriangleleft="3\@xm4A
\mathchardef\vartriangle="0\@xm4D
\mathchardef\blacktriangle="0\@xm4E
\mathchardef\triangledown="0\@xm4F
\mathchardef\eqcirc="3\@xm50
\mathchardef\lesseqgtr="3\@xm51
\mathchardef\gtreqless="3\@xm52
\mathchardef\lesseqqgtr="3\@xm53
\mathchardef\gtreqqless="3\@xm54
\mathchardef\Rrightarrow="3\@xm56
\mathchardef\Lleftarrow="3\@xm57
\mathchardef\veebar="2\@xm59
\mathchardef\barwedge="2\@xm5A
\mathchardef\doublebarwedge="2\@xm5B
\mathchardef\angle="0\@xm5C
\mathchardef\measuredangle="0\@xm5D
\mathchardef\sphericalangle="0\@xm5E
\mathchardef\varpropto="3\@xm5F
\mathchardef\smallsmile="3\@xm60
\mathchardef\smallfrown="3\@xm61
\mathchardef\Subset="3\@xm62
\mathchardef\Supset="3\@xm63
\mathchardef\Cup="2\@xm64

\mathchardef\Cap="2\@xm65

\mathchardef\curlywedge="2\@xm66
\mathchardef\curlyvee="2\@xm67
\mathchardef\leftthreetimes="2\@xm68
\mathchardef\rightthreetimes="2\@xm69
\mathchardef\subseteqq="3\@xm6A
\mathchardef\supseteqq="3\@xm6B
\mathchardef\bumpeq="3\@xm6C
\mathchardef\Bumpeq="3\@xm6D
\mathchardef\lll="3\@xm6E

\mathchardef\ggg="3\@xm6F

\mathchardef\circledS="0\@xm73
\mathchardef\pitchfork="3\@xm74
\mathchardef\dotplus="2\@xm75
\mathchardef\backsim="3\@xm76
\mathchardef\backsimeq="3\@xm77
\mathchardef\complement="0\@xm7B
\mathchardef\intercal="2\@xm7C
\mathchardef\circledcirc="2\@xm7D
\mathchardef\circledast="2\@xm7E
\mathchardef\circleddash="2\@xm7F
\def\ulcorner{\delimiter"4\@xm70\@xm70 }
\def\urcorner{\delimiter"5\@xm71\@xm71 }
\def\llcorner{\delimiter"4\@xm78\@xm78 }
\def\lrcorner{\delimiter"5\@xm79\@xm79 }
\def\yen{\mathhexbox\@xm55 }
\def\checkmark{\mathhexbox\@xm58 }
\def\circledR{\mathhexbox\@xm72 }
\def\maltese{\mathhexbox\@xm7A }
\mathchardef\lvertneqq="3\@ym00
\mathchardef\gvertneqq="3\@ym01
\mathchardef\nleq="3\@ym02
\mathchardef\ngeq="3\@ym03
\mathchardef\nless="3\@ym04
\mathchardef\ngtr="3\@ym05
\mathchardef\nprec="3\@ym06
\mathchardef\nsucc="3\@ym07
\mathchardef\lneqq="3\@ym08
\mathchardef\gneqq="3\@ym09
\mathchardef\nleqslant="3\@ym0A
\mathchardef\ngeqslant="3\@ym0B
\mathchardef\lneq="3\@ym0C
\mathchardef\gneq="3\@ym0D
\mathchardef\npreceq="3\@ym0E
\mathchardef\nsucceq="3\@ym0F
\mathchardef\precnsim="3\@ym10
\mathchardef\succnsim="3\@ym11
\mathchardef\lnsim="3\@ym12
\mathchardef\gnsim="3\@ym13
\mathchardef\nleqq="3\@ym14
\mathchardef\ngeqq="3\@ym15
\mathchardef\precneqq="3\@ym16
\mathchardef\succneqq="3\@ym17
\mathchardef\precnapprox="3\@ym18
\mathchardef\succnapprox="3\@ym19
\mathchardef\lnapprox="3\@ym1A
\mathchardef\gnapprox="3\@ym1B
\mathchardef\nsim="3\@ym1C
\mathchardef\ncong="3\@ym1D

\mathchardef\varsubsetneq="3\@ym20
\mathchardef\varsupsetneq="3\@ym21
\mathchardef\nsubseteqq="3\@ym22
\mathchardef\nsupseteqq="3\@ym23
\mathchardef\subsetneqq="3\@ym24
\mathchardef\supsetneqq="3\@ym25
\mathchardef\varsubsetneqq="3\@ym26
\mathchardef\varsupsetneqq="3\@ym27
\mathchardef\subsetneq="3\@ym28
\mathchardef\supsetneq="3\@ym29
\mathchardef\nsubseteq="3\@ym2A
\mathchardef\nsupseteq="3\@ym2B
\mathchardef\nparallel="3\@ym2C
\mathchardef\nmid="3\@ym2D
\mathchardef\nshortmid="3\@ym2E
\mathchardef\nshortparallel="3\@ym2F
\mathchardef\nvdash="3\@ym30
\mathchardef\nVdash="3\@ym31
\mathchardef\nvDash="3\@ym32
\mathchardef\nVDash="3\@ym33
\mathchardef\ntrianglerighteq="3\@ym34
\mathchardef\ntrianglelefteq="3\@ym35
\mathchardef\ntriangleleft="3\@ym36
\mathchardef\ntriangleright="3\@ym37
\mathchardef\nleftarrow="3\@ym38
\mathchardef\nrightarrow="3\@ym39
\mathchardef\nLeftarrow="3\@ym3A
\mathchardef\nRightarrow="3\@ym3B
\mathchardef\nLeftrightarrow="3\@ym3C
\mathchardef\nleftrightarrow="3\@ym3D
\mathchardef\divideontimes="2\@ym3E
\mathchardef\varnothing="0\@ym3F
\mathchardef\nexists="0\@ym40
\mathchardef\mho="0\@ym66
\mathchardef\eth="0\@ym67
\mathchardef\eqsim="3\@ym68
\mathchardef\beth="0\@ym69
\mathchardef\gimel="0\@ym6A
\mathchardef\daleth="0\@ym6B
\mathchardef\lessdot="3\@ym6C
\mathchardef\gtrdot="3\@ym6D
\mathchardef\ltimes="2\@ym6E
\mathchardef\rtimes="2\@ym6F
\mathchardef\shortmid="3\@ym70
\mathchardef\shortparallel="3\@ym71
\mathchardef\smallsetminus="2\@ym72
\mathchardef\thicksim="3\@ym73
\mathchardef\thickapprox="3\@ym74
\mathchardef\approxeq="3\@ym75
\mathchardef\succapprox="3\@ym76
\mathchardef\precapprox="3\@ym77
\mathchardef\curvearrowleft="3\@ym78
\mathchardef\curvearrowright="3\@ym79
\mathchardef\digamma="0\@ym7A
\mathchardef\varkappa="0\@ym7B
\mathchardef\hslash="0\@ym7D
\mathchardef\hbar="0\@ym7E
\mathchardef\backepsilon="3\@ym7F


\def\Bbb{\ifmmode\let\next\Bbb@\else
\def\next{\errmessage{Use \string\Bbb\space only in math mode}}\fi\next}
\def\Bbb@#1{{\Bbb@@{#1}}}
\def\Bbb@@#1{\fam\ymfam#1}
\fi


\def\Nulle{0} 
\def\Afe{1}   
\def\Hae{2}   
\def\Hbe{3}   
\def\Hce{4}   
\def\Hde{5}   


\newcount\LastMac       \LastMac=\Nulle

\newskip\half      \half=5.5pt plus 1.5pt minus 2.25pt
\newskip\one       \one=11pt plus 3pt minus 5.5pt
\newskip\onehalf   \onehalf=16.5pt plus 5.5pt minus 8.25pt
\newskip\two       \two=22pt plus 5.5pt minus 11pt

\def\Half{\addvspace{\half}}
\def\One{\addvspace{\one}}
\def\OneHalf{\addvspace{\onehalf}}
\def\Two{\addvspace{\two}}


\def\Raggedright{
  \rightskip=\z@ plus \hsize\relax
}

\def\Fullout{
  \rightskip=\z@\relax
}

\def\Hang#1#2{
  \hangindent=#1%
  \hangafter=#2\relax
}


\newif\ifsp@page
\def\pagestyle#1{\csname ps@#1\endcsname}
\def\thispagestyle#1{\global\sp@pagetrue\gdef\sp@type{#1}}

\def\ps@titlepage{%
  \def\@oddhead{\eightpoint\noindent \the\CatchLine
    \ifprod@font\else\qquad Printed\ \today\fi \hfil}%
  \let\@evenhead=\@oddhead
}

\def\ps@headings{%
  \def\@oddhead{\elevenpoint\it\noindent
    \hfill\the\RightHeader\hskip1.5em\rm\folio}%
  \def\@evenhead{\elevenpoint\noindent
    \folio\hskip1.5em\it\the\LeftHeader\hfill}%
}

\def\ps@plate{%
  \def\@oddhead{\eightpoint\noindent\plt@cap\hfil}%
  \def\@evenhead{\eightpoint\noindent\plt@cap\hfil}%
}



\def\title#1{
  \bgroup
    \vbox to 8pt{\vss}%
    \seventeenpoint
    \Raggedright
    \noindent \strut{\bf #1}\par
  \egroup
}

\def\author#1{
  \bgroup
    \ifnum\LastMac=\Afe \OneHalf\else \vskip 21pt\fi
    \fourteenpoint
    \Raggedright
    \noindent \strut #1\par
    \vskip 3pt%
  \egroup
}

\def\affiliation#1{
  \bgroup
    \vskip -4pt%
    \eightpoint
    \Raggedright
    \noindent \strut {\it #1}\par
  \egroup
  \LastMac=\Afe\relax
}

\def\acceptedline#1{
  \bgroup
    \Two
    \eightpoint
    \Raggedright
    \noindent \strut #1\par
  \egroup
}

\long\def\abstract#1{%
  \bgroup
    \vskip 20pt%
    \everypar{\Hang{11pc}{0}}%
    \noindent{\ninebf ABSTRACT}\par
    \tenpoint
    \Fullout
    \noindent #1\par
  \egroup
}

\long\def\keywords#1{
  \bgroup
    \Half
    \everypar{\Hang{11pc}{0}}%
    \tenpoint
    \Fullout
    \noindent\hbox{\bf Key words:}\ #1\par
  \egroup
}


\def\maketitle{%
  \EndOpening
  \ifsinglecol \else \MakePage\fi
}


\def\pageoffset#1#2{\hoffset=#1\relax\voffset=#2\relax}


\def\Autonumber{
  \global\AutoNumbertrue  
}

\newif\ifAutoNumber \AutoNumberfalse
\newcount\Sec        
\newcount\SecSec
\newcount\SecSecSec

\Sec=\z@

\def\:{\let\@sptoken= } \:  
\def\:{\@xifnch} \expandafter\def\: {\futurelet\@tempc\@ifnch}

\def\@ifnextchar#1#2#3{%
  \let\@tempMACe #1%
  \def\@tempMACa{#2}%
  \def\@tempMACb{#3}%
  \futurelet \@tempMACc\@ifnch%
}

\def\@ifnch{%
\ifx \@tempMACc \@sptoken%
  \let\@tempMACd\@xifnch%
\else%
  \ifx \@tempMACc \@tempMACe%
    \let\@tempMACd\@tempMACa%
  \else%
    \let\@tempMACd\@tempMACb%
  \fi%
\fi%
\@tempMACd%
}

\def\@ifstar#1#2{\@ifnextchar *{\def\@tempMACa*{#1}\@tempMACa}{#2}}

\newskip\@tempskipb

\def\addvspace#1{%
  \ifvmode\else \endgraf\fi%
  \ifdim\lastskip=\z@%
    \vskip #1\relax%
  \else%
    \@tempskipb#1\relax\@xaddvskip%
  \fi%
}

\def\@xaddvskip{%
  \ifdim\lastskip<\@tempskipb%
    \vskip-\lastskip%
    \vskip\@tempskipb\relax%
  \else%
    \ifdim\@tempskipb<\z@%
      \ifdim\lastskip<\z@ \else%
        \advance\@tempskipb\lastskip%
        \vskip-\lastskip\vskip\@tempskipb%
      \fi%
    \fi%
  \fi%
}

\newskip\@tmpSKIP

\def\addpen#1{%
  \ifvmode
    \if@nobreak
    \else
      \ifdim\lastskip=\z@
        \penalty#1\relax
      \else
        \@tmpSKIP=\lastskip
        \vskip -\lastskip
        \penalty#1\vskip\@tmpSKIP
      \fi
    \fi
  \fi
}

\newcount\@clubpen   \@clubpen=\clubpenalty
\newif\if@nobreak    \@nobreakfalse

\def\@noafterindent{%
  \global\@nobreaktrue
  \everypar{\if@nobreak
              \global\@nobreakfalse
              \clubpenalty \@M
              {\setbox\z@\lastbox}%
              \LastMac=\Nulle\relax%
            \else
              \clubpenalty \@clubpen
              \everypar{}%
            \fi}
}

\newcount\gds@cbrk   \gds@cbrk=-300

\def\@nohdbrk{\interlinepenalty \@M\relax}

\let\@par=\par
\def\@restorepar{\def\par{\@par}}

\newif\if@endpe   \@endpefalse

\def\@doendpe{\@endpetrue \@nobreakfalse \LastMac=\Nulle\relax%
     \def\par{\@restorepar\everypar{}\par\@endpefalse}%
              \everypar{\setbox\z@\lastbox\everypar{}\@endpefalse}%
}

\def\section{\@ifstar{\@ssection}{\@section}}

\def\@section#1{
  \if@nobreak
    \everypar{}%
    \ifnum\LastMac=\Hae \addvspace{\half}\fi
  \else
    \addpen{\gds@cbrk}%
    \addvspace{\two}%
  \fi
  \bgroup
    \ninepoint\bf
    \Raggedright
    \ifAutoNumber
      \global\advance\Sec \@ne
      \noindent\@nohdbrk\number\Sec\hskip 1pc \uppercase{#1}\par
      \global\SecSec=\z@
    \else
      \noindent\@nohdbrk\uppercase{#1}\par
    \fi
  \egroup
  \nobreak
  \vskip\half
  \nobreak
  \@noafterindent
  \LastMac=\Hae\relax
}

\def\@ssection#1{
  \if@nobreak
    \everypar{}%
    \ifnum\LastMac=\Hae \addvspace{\half}\fi
  \else
    \addpen{\gds@cbrk}%
    \addvspace{\two}%
  \fi
  \bgroup
    \ninepoint\bf
    \Raggedright
    \noindent\@nohdbrk\uppercase{#1}\par
  \egroup
  \nobreak
  \vskip\half
  \nobreak
  \@noafterindent
  \LastMac=\Hae\relax
}

\def\subsection#1{
  \if@nobreak
    \everypar{}%
    \ifnum\LastMac=\Hae \addvspace{1pt plus 1pt minus .5pt}\fi
  \else
    \addpen{\gds@cbrk}%
    \addvspace{\onehalf}%
  \fi
  \bgroup
    \ninepoint\bf
    \Raggedright
    \ifAutoNumber
      \global\advance\SecSec \@ne
      \noindent\@nohdbrk\number\Sec.\number\SecSec \hskip 1pc\relax #1\par
      \global\SecSecSec=\z@
    \else
      \noindent\@nohdbrk #1\par
    \fi
  \egroup
  \nobreak
  \vskip\half
  \nobreak
  \@noafterindent
  \LastMac=\Hbe\relax
}

\def\subsubsection#1{
  \if@nobreak
    \everypar{}%
    \ifnum\LastMac=\Hbe \addvspace{1pt plus 1pt minus .5pt}\fi
  \else
    \addpen{\gds@cbrk}%
    \addvspace{\onehalf}%
  \fi
  \bgroup
    \ninepoint\it
    \Raggedright
    \ifAutoNumber
      \global\advance\SecSecSec \@ne
      \noindent\@nohdbrk\number\Sec.\number\SecSec.\number\SecSecSec
        \hskip 1pc\relax #1\par
    \else
      \noindent\@nohdbrk #1\par
    \fi
  \egroup
  \nobreak
  \vskip\half
  \nobreak
  \@noafterindent
  \LastMac=\Hce\relax
}

\def\paragraph#1{
  \if@nobreak
    \everypar{}%
  \else
    \addpen{\gds@cbrk}%
    \addvspace{\one}%
  \fi%
  \bgroup%
    \ninepoint\it
    \noindent #1\ \nobreak%
  \egroup
  \LastMac=\Hde\relax
  \ignorespaces
}




\def\beginlist{%
  \par\if@nobreak \else\addvspace{\half}\fi%
  \bgroup%
    \ninepoint
    \let\item=\list@item%
}

\def\list@item{%
  \par\noindent\hskip 1em\relax%
  \ignorespaces%
}

\def\endlist{\par\egroup\addvspace{\half}\@doendpe}


\def\beginrefs{%
  \par
  \bgroup
    \eightpoint
    \Raggedright
    \let\bibitem=\bib@item
}

\def\bib@item{%
  \par\parindent=1.5em\Hang{1.5em}{1}%
  \everypar={\Hang{1.5em}{1}\ignorespaces}%
  \noindent\ignorespaces
}

\def\endrefs{\par\egroup\@doendpe}


\newtoks\CatchLine

\def\@journal{Mon.\ Not.\ R.\ Astron.\ Soc.\ }  
\def\@pubyear{1994}        
\def\@pagerange{000--000}  
\def\@volume{000}          
\def\@microfiche{}         %

\def\pubyear#1{\gdef\@pubyear{#1}\@makecatchline}
\def\pagerange#1{\gdef\@pagerange{#1}\@makecatchline}
\def\volume#1{\gdef\@volume{#1}\@makecatchline}
\def\microfiche#1{\gdef\@microfiche{and Microfiche\ #1}\@makecatchline}

\def\@makecatchline{%
  \global\CatchLine{%
    {\rm \@journal {\bf \@volume},\ \@pagerange\ (\@pubyear)\ \@microfiche}}%
}

\@makecatchline 

\newtoks\LeftHeader
\def\shortauthor#1{
  \global\LeftHeader{#1}%
}

\newtoks\RightHeader
\def\shorttitle#1{
  \global\RightHeader{#1}%
}

\def\PageHead{
  \begingroup
    \ifsp@page
      \csname ps@\sp@type\endcsname
      \global\sp@pagefalse
    \fi
    \ifodd\pageno
      \let\the@head=\@oddhead
    \else
      \let\the@head=\@evenhead
    \fi
    \vbox to \z@{\vskip-22.5\p@%
      \hbox to \PageWidth{\vbox to8.5\p@{}%
        \the@head
      }%
    \vss}%
  \endgroup
  \nointerlineskip
}

\def\today{%
  \number\day\space
  \ifcase\month\or January\or February\or March\or April\or May\or June\or
    July\or August\or September\or October\or November\or December\fi
  \space\number\year%
}

\def\PageFoot{} 

\def\authorcomment#1{%
  \gdef\PageFoot{%
    \nointerlineskip%
    \vbox to 22pt{\vfil%
      \hbox to \PageWidth{\elevenpoint\noindent \hfil #1 \hfil}}%
  }%
}


\newif\ifplate@page
\newbox\plt@box

\def\beginplatepage{%
  \let\plate=\plate@head
  \let\caption=\fig@caption
  \global\setbox\plt@box=\vbox\bgroup
  \TEMPDIMEN=\PageWidth 
  \hsize=\PageWidth\relax
}

\def\endplatepage{\par\egroup\global\plate@pagetrue}
\def\plate@head#1{\gdef\plt@cap{#1}}


\def\letters{%
  \gdef\folio{\ifnum\pageno<\z@ L\romannumeral-\pageno
    \else L\number\pageno \fi}%
}


\everydisplay{\displaysetup}

\newif\ifeqno
\newif\ifleqno

\def\displaysetup#1$${%
 \displaytest#1\eqno\eqno\displaytest
}

\def\displaytest#1\eqno#2\eqno#3\displaytest{%
 \if!#3!\ldisplaytest#1\leqno\leqno\ldisplaytest
 \else\eqnotrue\leqnofalse\def\eqn{#2}\def\eq{#1}\fi
 \generaldisplay$$}

\def\ldisplaytest#1\leqno#2\leqno#3\ldisplaytest{%
 \def\eq{#1}%
 \if!#3!\eqnofalse\else\eqnotrue\leqnotrue
  \def\eqn{#2}\fi}

\def\generaldisplay{%
\ifeqno \ifleqno
   \hbox to \hsize{\noindent
     $\displaystyle\eq$\hfil$\displaystyle\eqn$}
  \else
    \hbox to \hsize{\noindent
     $\displaystyle\eq$\hfil$\displaystyle\eqn$}
  \fi
 \else
 \hbox to \hsize{\vbox{\noindent
  $\displaystyle\eq$\hfil}}
 \fi
}


\def\@notice{%
  \par\Two%
  \noindent{\b@ls{11pt}\ninerm This paper has been produced using the
    Blackwell Scientific Publications \TeX\ macros.\par}%
}

\outer\def\bye{\@notice\par\vfill\supereject\end}


\def\start@mess{%
  Monthly notices of the RAS journal style (\@typeface)\space
    v\@version,\space \@verdate.%
}

\everyjob{\Warn{\start@mess}}



\newif\if@debug \@debugfalse  

\def\Print#1{\if@debug\immediate\write16{#1}\else \fi}
\def\Warn#1{\immediate\write16{#1}}
\def\wlog#1{}

\newcount\Iteration 

\def\Single{0} \def\Double{1}                 
\def\Figure{0} \def\Table{1}                  

\def\InStack{0}  
\def\InZoneA{1}
\def\InZoneB{2}
\def\InZoneC{3}

\newcount\TEMPCOUNT 
\newdimen\TEMPDIMEN 
\newbox\TEMPBOX     
\newbox\VOIDBOX     

\newcount\LengthOfStack 
\newcount\MaxItems      
\newcount\StackPointer
\newcount\Point         
\newcount\NextFigure    
\newcount\NextTable     
\newcount\NextItem      

\newcount\StatusStack   
\newcount\NumStack      
\newcount\TypeStack     
\newcount\SpanStack     
\newcount\BoxStack      

\newcount\ItemSTATUS    
\newcount\ItemNUMBER    
\newcount\ItemTYPE      
\newcount\ItemSPAN      
\newbox\ItemBOX         
\newdimen\ItemSIZE      

\newdimen\PageHeight    
\newdimen\TextLeading   
\newdimen\Feathering    
\newcount\LinesPerPage  
\newdimen\ColumnWidth   
\newdimen\ColumnGap     
\newdimen\PageWidth     
\newdimen\BodgeHeight   
\newcount\Leading       

\newdimen\ZoneBSize  
\newdimen\TextSize   
\newbox\ZoneABOX     
\newbox\ZoneBBOX     
\newbox\ZoneCBOX     

\newif\ifFirstSingleItem
\newif\ifFirstZoneA
\newif\ifMakePageInComplete
\newif\ifMoreFigures \MoreFiguresfalse 
\newif\ifMoreTables  \MoreTablesfalse  

\newif\ifFigInZoneB 
\newif\ifFigInZoneC 
\newif\ifTabInZoneB 
\newif\ifTabInZoneC

\newif\ifZoneAFullPage

\newbox\MidBOX    
\newbox\LeftBOX
\newbox\RightBOX
\newbox\PageBOX   

\newif\ifLeftCOL  
\LeftCOLtrue

\newdimen\ZoneBAdjust

\newcount\ItemFits
\def\Yes{1}
\def\No{2}


\MaxItems=15
\NextFigure=\z@        
\NextTable=\@ne

\BodgeHeight=6pt
\TextLeading=11pt    
\Leading=11
\Feathering=\z@      
\LinesPerPage=61     
\topskip=\TextLeading
\ColumnWidth=20pc    
\ColumnGap=2pc       

\newskip\ItemSepamount  
\ItemSepamount=\TextLeading plus \TextLeading minus 4pt

\parskip=\z@ plus .1pt
\parindent=18pt
\widowpenalty=\z@
\clubpenalty=10000
\tolerance=1500
\hbadness=1500
\abovedisplayskip=6pt plus 2pt minus 2pt
\belowdisplayskip=6pt plus 2pt minus 2pt
\abovedisplayshortskip=6pt plus 2pt minus 2pt
\belowdisplayshortskip=6pt plus 2pt minus 2pt

\ninepoint 


\PageHeight=682pt

\PageWidth=2\ColumnWidth
\advance\PageWidth by \ColumnGap

\pagestyle{headings}




\newcount\DUMMY \StatusStack=\allocationnumber
\newcount\DUMMY \newcount\DUMMY \newcount\DUMMY
\newcount\DUMMY \newcount\DUMMY \newcount\DUMMY
\newcount\DUMMY \newcount\DUMMY \newcount\DUMMY
\newcount\DUMMY \newcount\DUMMY \newcount\DUMMY
\newcount\DUMMY \newcount\DUMMY \newcount\DUMMY

\newcount\DUMMY \NumStack=\allocationnumber
\newcount\DUMMY \newcount\DUMMY \newcount\DUMMY
\newcount\DUMMY \newcount\DUMMY \newcount\DUMMY
\newcount\DUMMY \newcount\DUMMY \newcount\DUMMY
\newcount\DUMMY \newcount\DUMMY \newcount\DUMMY
\newcount\DUMMY \newcount\DUMMY \newcount\DUMMY

\newcount\DUMMY \TypeStack=\allocationnumber
\newcount\DUMMY \newcount\DUMMY \newcount\DUMMY
\newcount\DUMMY \newcount\DUMMY \newcount\DUMMY
\newcount\DUMMY \newcount\DUMMY \newcount\DUMMY
\newcount\DUMMY \newcount\DUMMY \newcount\DUMMY
\newcount\DUMMY \newcount\DUMMY \newcount\DUMMY

\newcount\DUMMY \SpanStack=\allocationnumber
\newcount\DUMMY \newcount\DUMMY \newcount\DUMMY
\newcount\DUMMY \newcount\DUMMY \newcount\DUMMY
\newcount\DUMMY \newcount\DUMMY \newcount\DUMMY
\newcount\DUMMY \newcount\DUMMY \newcount\DUMMY
\newcount\DUMMY \newcount\DUMMY \newcount\DUMMY

\newbox\DUMMY   \BoxStack=\allocationnumber
\newbox\DUMMY   \newbox\DUMMY \newbox\DUMMY
\newbox\DUMMY   \newbox\DUMMY \newbox\DUMMY
\newbox\DUMMY   \newbox\DUMMY \newbox\DUMMY
\newbox\DUMMY   \newbox\DUMMY \newbox\DUMMY
\newbox\DUMMY   \newbox\DUMMY \newbox\DUMMY

\def\wlog{\immediate\write\m@ne}


\def\GetItemAll#1{%
 \GetItemSTATUS{#1}
 \GetItemNUMBER{#1}
 \GetItemTYPE{#1}
 \GetItemSPAN{#1}
 \GetItemBOX{#1}
}

\def\GetItemSTATUS#1{%
 \Point=\StatusStack
 \advance\Point by #1
 \global\ItemSTATUS=\count\Point
}

\def\GetItemNUMBER#1{%
 \Point=\NumStack
 \advance\Point by #1
 \global\ItemNUMBER=\count\Point
}

\def\GetItemTYPE#1{%
 \Point=\TypeStack
 \advance\Point by #1
 \global\ItemTYPE=\count\Point
}

\def\GetItemSPAN#1{%
 \Point\SpanStack
 \advance\Point by #1
 \global\ItemSPAN=\count\Point
}

\def\GetItemBOX#1{%
 \Point=\BoxStack
 \advance\Point by #1
 \global\setbox\ItemBOX=\vbox{\copy\Point}
 \global\ItemSIZE=\ht\ItemBOX
 \global\advance\ItemSIZE by \dp\ItemBOX
 \TEMPCOUNT=\ItemSIZE
 \divide\TEMPCOUNT by \Leading
 \divide\TEMPCOUNT by 65536
 \advance\TEMPCOUNT \@ne
 \ItemSIZE=\TEMPCOUNT pt
 \global\multiply\ItemSIZE by \Leading
}


\def\JoinStack{%
 \ifnum\LengthOfStack=\MaxItems 
  \Warn{WARNING: Stack is full...some items will be lost!}
 \else
  \Point=\StatusStack
  \advance\Point by \LengthOfStack
  \global\count\Point=\ItemSTATUS
  \Point=\NumStack
  \advance\Point by \LengthOfStack
  \global\count\Point=\ItemNUMBER
  \Point=\TypeStack
  \advance\Point by \LengthOfStack
  \global\count\Point=\ItemTYPE
  \Point\SpanStack
  \advance\Point by \LengthOfStack
  \global\count\Point=\ItemSPAN
  \Point=\BoxStack
  \advance\Point by \LengthOfStack
  \global\setbox\Point=\vbox{\copy\ItemBOX}
  \global\advance\LengthOfStack \@ne
  \ifnum\ItemTYPE=\Figure 
   \global\MoreFigurestrue
  \else
   \global\MoreTablestrue
  \fi
 \fi
}


\def\LeaveStack#1{%
 {\Iteration=#1
 \loop
 \ifnum\Iteration<\LengthOfStack
  \advance\Iteration \@ne
  \GetItemSTATUS{\Iteration}
   \advance\Point by \m@ne
   \global\count\Point=\ItemSTATUS
  \GetItemNUMBER{\Iteration}
   \advance\Point by \m@ne
   \global\count\Point=\ItemNUMBER
  \GetItemTYPE{\Iteration}
   \advance\Point by \m@ne
   \global\count\Point=\ItemTYPE
  \GetItemSPAN{\Iteration}
   \advance\Point by \m@ne
   \global\count\Point=\ItemSPAN
  \GetItemBOX{\Iteration}
   \advance\Point by \m@ne
   \global\setbox\Point=\vbox{\copy\ItemBOX}
 \repeat}
 \global\advance\LengthOfStack by \m@ne
}


\newif\ifStackNotClean

\def\CleanStack{%
 \StackNotCleantrue
 {\Iteration=\z@
  \loop
   \ifStackNotClean
    \GetItemSTATUS{\Iteration}
    \ifnum\ItemSTATUS=\InStack
     \advance\Iteration \@ne
     \else
      \LeaveStack{\Iteration}
    \fi
   \ifnum\LengthOfStack<\Iteration
    \StackNotCleanfalse
   \fi
 \repeat}
}


\def\FindItem#1#2{%
 \global\StackPointer=\m@ne 
 {\Iteration=\z@
  \loop
  \ifnum\Iteration<\LengthOfStack
   \GetItemSTATUS{\Iteration}
   \ifnum\ItemSTATUS=\InStack
    \GetItemTYPE{\Iteration}
    \ifnum\ItemTYPE=#1
     \GetItemNUMBER{\Iteration}
     \ifnum\ItemNUMBER=#2
      \global\StackPointer=\Iteration
      \Iteration=\LengthOfStack 
     \fi
    \fi
   \fi
  \advance\Iteration \@ne
 \repeat}
}


\def\FindNext{%
 \global\StackPointer=\m@ne 
 {\Iteration=\z@
  \loop
  \ifnum\Iteration<\LengthOfStack
   \GetItemSTATUS{\Iteration}
   \ifnum\ItemSTATUS=\InStack
    \GetItemTYPE{\Iteration}
   \ifnum\ItemTYPE=\Figure
    \ifMoreFigures
      \global\NextItem=\Figure
      \global\StackPointer=\Iteration
      \Iteration=\LengthOfStack 
    \fi
   \fi
   \ifnum\ItemTYPE=\Table
    \ifMoreTables
      \global\NextItem=\Table
      \global\StackPointer=\Iteration
      \Iteration=\LengthOfStack 
    \fi
   \fi
  \fi
  \advance\Iteration \@ne
 \repeat}
}


\def\ChangeStatus#1#2{%
 \Point=\StatusStack
 \advance\Point by #1
 \global\count\Point=#2
}



\def\Zone{\InZoneA}

\ZoneBAdjust=\z@

\def\MakePage{
 \global\ZoneBSize=\PageHeight
 \global\TextSize=\ZoneBSize
 \global\ZoneAFullPagefalse
 \global\topskip=\TextLeading
 \MakePageInCompletetrue
 \MoreFigurestrue
 \MoreTablestrue
 \FigInZoneBfalse
 \FigInZoneCfalse
 \TabInZoneBfalse
 \TabInZoneCfalse
 \global\FirstSingleItemtrue
 \global\FirstZoneAtrue
 \global\setbox\ZoneABOX=\box\VOIDBOX
 \global\setbox\ZoneBBOX=\box\VOIDBOX
 \global\setbox\ZoneCBOX=\box\VOIDBOX
 \loop
  \ifMakePageInComplete
 \FindNext
 \ifnum\StackPointer=\m@ne
  \NextItem=\m@ne
  \MoreFiguresfalse
  \MoreTablesfalse
 \fi
 \ifnum\NextItem=\Figure
   \FindItem{\Figure}{\NextFigure}
   \ifnum\StackPointer=\m@ne \global\MoreFiguresfalse
   \else
    \GetItemSPAN{\StackPointer}
    \ifnum\ItemSPAN=\Single \def\Zone{\InZoneB}\relax
     \ifFigInZoneC \global\MoreFiguresfalse\fi
    \else
     \def\Zone{\InZoneA}
     \ifFigInZoneB \def\Zone{\InZoneC}\fi
    \fi
   \fi
   \ifMoreFigures\Print{}\FigureItems\fi
 \fi
\ifnum\NextItem=\Table
   \FindItem{\Table}{\NextTable}
   \ifnum\StackPointer=\m@ne \global\MoreTablesfalse
   \else
    \GetItemSPAN{\StackPointer}
    \ifnum\ItemSPAN=\Single\relax
     \ifTabInZoneC \global\MoreTablesfalse\fi
    \else
     \def\Zone{\InZoneA}
     \ifTabInZoneB \def\Zone{\InZoneC}\fi
    \fi
   \fi
   \ifMoreTables\Print{}\TableItems\fi
 \fi
   \MakePageInCompletefalse 
   \ifMoreFigures\MakePageInCompletetrue\fi
   \ifMoreTables\MakePageInCompletetrue\fi
 \repeat
 \ifZoneAFullPage
  \global\TextSize=\z@
  \global\ZoneBSize=\z@
  \global\vsize=\z@\relax
  \global\topskip=\z@\relax
  \vbox to \z@{\vss}
  \eject
 \else
 \global\advance\ZoneBSize by -\ZoneBAdjust
 \global\vsize=\ZoneBSize
 \global\hsize=\ColumnWidth
 \global\ZoneBAdjust=\z@
 \ifdim\TextSize<23pt
 \Warn{}
 \Warn{* Making column fall short: TextSize=\the\TextSize *}
 \vskip-\lastskip\eject\fi
 \fi
}

\def\MakeRightCol{
 \global\TextSize=\ZoneBSize
 \MakePageInCompletetrue
 \MoreFigurestrue
 \MoreTablestrue
 \global\FirstSingleItemtrue
 \global\setbox\ZoneBBOX=\box\VOIDBOX
 \def\Zone{\InZoneB}
 \loop
  \ifMakePageInComplete
 \FindNext
 \ifnum\StackPointer=\m@ne
  \NextItem=\m@ne
  \MoreFiguresfalse
  \MoreTablesfalse
 \fi
 \ifnum\NextItem=\Figure
   \FindItem{\Figure}{\NextFigure}
   \ifnum\StackPointer=\m@ne \MoreFiguresfalse
   \else
    \GetItemSPAN{\StackPointer}
    \ifnum\ItemSPAN=\Double\relax
     \MoreFiguresfalse\fi
   \fi
   \ifMoreFigures\Print{}\FigureItems\fi
 \fi
 \ifnum\NextItem=\Table
   \FindItem{\Table}{\NextTable}
   \ifnum\StackPointer=\m@ne \MoreTablesfalse
   \else
    \GetItemSPAN{\StackPointer}
    \ifnum\ItemSPAN=\Double\relax
     \MoreTablesfalse\fi
   \fi
   \ifMoreTables\Print{}\TableItems\fi
 \fi
   \MakePageInCompletefalse 
   \ifMoreFigures\MakePageInCompletetrue\fi
   \ifMoreTables\MakePageInCompletetrue\fi
 \repeat
 \ifZoneAFullPage
  \global\TextSize=\z@
  \global\ZoneBSize=\z@
  \global\vsize=\z@\relax
  \global\topskip=\z@\relax
  \vbox to \z@{\vss}
  \eject
 \else
 \global\vsize=\ZoneBSize
 \global\hsize=\ColumnWidth
 \ifdim\TextSize<23pt
 \Warn{}
 \Warn{* Making column fall short: TextSize=\the\TextSize *}
 \vskip-\lastskip\eject\fi
\fi
}

\def\FigureItems{
 \Print{Considering...}
 \ShowItem{\StackPointer}
 \GetItemBOX{\StackPointer} 
 \GetItemSPAN{\StackPointer}
  \CheckFitInZone 
  \ifnum\ItemFits=\Yes
   \ifnum\ItemSPAN=\Single
     \ChangeStatus{\StackPointer}{\InZoneB} 
     \global\FigInZoneBtrue
     \ifFirstSingleItem
      \hbox{}\vskip-\BodgeHeight
     \global\advance\ItemSIZE by \TextLeading
     \fi
     \unvbox\ItemBOX\ItemSep
     \global\FirstSingleItemfalse
     \global\advance\TextSize by -\ItemSIZE
     \global\advance\TextSize by -\TextLeading
   \else
    \ifFirstZoneA
     \global\advance\ItemSIZE by \TextLeading
     \global\FirstZoneAfalse\fi
    \global\advance\TextSize by -\ItemSIZE
    \global\advance\TextSize by -\TextLeading
    \global\advance\ZoneBSize by -\ItemSIZE
    \global\advance\ZoneBSize by -\TextLeading
    \ifFigInZoneB\relax
     \else
     \ifdim\TextSize<3\TextLeading
     \global\ZoneAFullPagetrue
     \fi
    \fi
    \ChangeStatus{\StackPointer}{\Zone}
    \ifnum\Zone=\InZoneC \global\FigInZoneCtrue\fi
  \fi
   \Print{TextSize=\the\TextSize}
   \Print{ZoneBSize=\the\ZoneBSize}
  \global\advance\NextFigure \@ne
   \Print{This figure has been placed.}
  \else
   \Print{No space available for this figure...holding over.}
   \Print{}
   \global\MoreFiguresfalse
  \fi
}

\def\TableItems{
 \Print{Considering...}
 \ShowItem{\StackPointer}
 \GetItemBOX{\StackPointer} 
 \GetItemSPAN{\StackPointer}
  \CheckFitInZone 
  \ifnum\ItemFits=\Yes
   \ifnum\ItemSPAN=\Single
    \ChangeStatus{\StackPointer}{\InZoneB}
     \global\TabInZoneBtrue
     \ifFirstSingleItem
      \hbox{}\vskip-\BodgeHeight
     \global\advance\ItemSIZE by \TextLeading
     \fi
     \unvbox\ItemBOX\ItemSep
     \global\FirstSingleItemfalse
     \global\advance\TextSize by -\ItemSIZE
     \global\advance\TextSize by -\TextLeading
   \else
    \ifFirstZoneA
    \global\advance\ItemSIZE by \TextLeading
    \global\FirstZoneAfalse\fi
    \global\advance\TextSize by -\ItemSIZE
    \global\advance\TextSize by -\TextLeading
    \global\advance\ZoneBSize by -\ItemSIZE
    \global\advance\ZoneBSize by -\TextLeading
    \ifFigInZoneB\relax
     \else
     \ifdim\TextSize<3\TextLeading
     \global\ZoneAFullPagetrue
     \fi
    \fi
    \ChangeStatus{\StackPointer}{\Zone}
    \ifnum\Zone=\InZoneC \global\TabInZoneCtrue\fi
   \fi
  \global\advance\NextTable \@ne
   \Print{This table has been placed.}
  \else
  \Print{No space available for this table...holding over.}
   \Print{}
   \global\MoreTablesfalse
  \fi
}


\def\CheckFitInZone{%
{\advance\TextSize by -\ItemSIZE
 \advance\TextSize by -\TextLeading
 \ifFirstSingleItem
  \advance\TextSize by \TextLeading
 \fi
 \ifnum\Zone=\InZoneA\relax
  \else \advance\TextSize by -\ZoneBAdjust
 \fi
 \ifdim\TextSize<3\TextLeading \global\ItemFits=\No
 \else \global\ItemFits=\Yes\fi}
}

\def\BeginOpening{%
  \thispagestyle{titlepage}%
  \global\setbox\ItemBOX=\vbox\bgroup%
    \hsize=\PageWidth%
    \hrule height \z@
    \ifsinglecol\vskip 6pt\fi 
}

\let\begintopmatter=\BeginOpening  

\def\EndOpening{%
  \One
  \egroup
  \ifsinglecol
    \box\ItemBOX%
    \vskip\TextLeading plus 2\TextLeading
    \@noafterindent
  \else
    \ItemNUMBER=\z@%
    \ItemTYPE=\Figure
    \ItemSPAN=\Double
    \ItemSTATUS=\InStack
    \JoinStack
  \fi
}


\newif\if@here  \@herefalse

\def\no@float{\global\@heretrue}
\let\nofloat=\relax 

\def\beginfigure{%
  \@ifstar{\global\@dfloattrue \@bfigure}{\global\@dfloatfalse \@bfigure}%
}

\def\@bfigure#1{%
  \par
  \if@dfloat
    \ItemSPAN=\Double
    \TEMPDIMEN=\PageWidth
  \else
    \ItemSPAN=\Single
    \TEMPDIMEN=\ColumnWidth
  \fi
  \ifsinglecol
    \TEMPDIMEN=\PageWidth
  \else
    \ItemSTATUS=\InStack
    \ItemNUMBER=#1%
    \ItemTYPE=\Figure
  \fi
  \bgroup
    \hsize=\TEMPDIMEN
    \global\setbox\ItemBOX=\vbox\bgroup
      \eightpoint\nostb@ls{10pt}%
      \let\caption=\fig@caption
      \ifsinglecol \let\nofloat=\no@float\fi
}

\def\fig@caption#1{%
  \vskip 5.5pt plus 6pt%
  \bgroup 
    \eightpoint\nostb@ls{10pt}%
    \setbox\TEMPBOX=\hbox{#1}%
    \ifdim\wd\TEMPBOX>\TEMPDIMEN
      \noindent \unhbox\TEMPBOX\par
    \else
      \hbox to \hsize{\hfil\unhbox\TEMPBOX\hfil}%
    \fi
  \egroup
}

\def\endfigure{%
  \par\egroup 
  \egroup
  \ifsinglecol
    \if@here \midinsert\global\@herefalse\else \topinsert\fi
      \unvbox\ItemBOX
    \endinsert
  \else
    \JoinStack
    \Print{Processing source for figure \the\ItemNUMBER}%
  \fi
}


\newbox\tab@cap@box
\def\tab@caption#1{\global\setbox\tab@cap@box=\hbox{#1\par}}

\newtoks\tab@txt@toks
\long\def\tab@txt#1{\global\tab@txt@toks={#1}\global\table@txttrue}

\newif\iftable@txt  \table@txtfalse
\newif\if@dfloat    \@dfloatfalse

\def\begintable{%
  \@ifstar{\global\@dfloattrue \@btable}{\global\@dfloatfalse \@btable}%
}

\def\@btable#1{%
  \par
  \if@dfloat
    \ItemSPAN=\Double
    \TEMPDIMEN=\PageWidth
  \else
    \ItemSPAN=\Single
    \TEMPDIMEN=\ColumnWidth
  \fi
  \ifsinglecol
    \TEMPDIMEN=\PageWidth
  \else
    \ItemSTATUS=\InStack
    \ItemNUMBER=#1%
    \ItemTYPE=\Table
  \fi
  \bgroup
    \eightpoint\nostb@ls{10pt}%
    \global\setbox\ItemBOX=\vbox\bgroup
      \let\caption=\tab@caption
      \let\tabletext=\tab@txt
      \ifsinglecol \let\nofloat=\no@float\fi
}

\def\endtable{%
  \par\egroup 
  \egroup
  \setbox\TEMPBOX=\hbox to \TEMPDIMEN{%
    \hss
    \vbox{%
      \hsize=\wd\ItemBOX
      \ifvoid\tab@cap@box
      \else
        \noindent\unhbox\tab@cap@box
        \vskip 5.5pt plus 6pt%
      \fi
      \box\ItemBOX
      \iftable@txt
        \vskip 10pt%
        \eightpoint\nostb@ls{10pt}%
        \noindent\the\tab@txt@toks
        \global\table@txtfalse
      \fi
    }%
    \hss
  }%
  \ifsinglecol
    \if@here \midinsert\global\@herefalse\else \topinsert\fi
      \box\TEMPBOX
    \endinsert
  \else
    \global\setbox\ItemBOX=\box\TEMPBOX
    \JoinStack
    \Print{Processing source for table \the\ItemNUMBER}%
  \fi
}

\def\UnloadZoneA{%
\FirstZoneAtrue
 \Iteration=\z@
  \loop
   \ifnum\Iteration<\LengthOfStack
    \GetItemSTATUS{\Iteration}
    \ifnum\ItemSTATUS=\InZoneA
     \GetItemBOX{\Iteration}
     \ifFirstZoneA \vbox to \BodgeHeight{\vfil}%
     \FirstZoneAfalse\fi
     \unvbox\ItemBOX\ItemSep
     \LeaveStack{\Iteration}
     \else
     \advance\Iteration \@ne
   \fi
 \repeat
}

\def\UnloadZoneC{%
\Iteration=\z@
  \loop
   \ifnum\Iteration<\LengthOfStack
    \GetItemSTATUS{\Iteration}
    \ifnum\ItemSTATUS=\InZoneC
     \GetItemBOX{\Iteration}
     \ItemSep\unvbox\ItemBOX
     \LeaveStack{\Iteration}
     \else
     \advance\Iteration \@ne
   \fi
 \repeat
}


\def\ShowItem#1{
  {\GetItemAll{#1}
  \Print{\the#1:
  {TYPE=\ifnum\ItemTYPE=\Figure Figure\else Table\fi}
  {NUMBER=\the\ItemNUMBER}
  {SPAN=\ifnum\ItemSPAN=\Single Single\else Double\fi}
  {SIZE=\the\ItemSIZE}}}
}

\def\ShowStack{%
 \Print{}
 \Print{LengthOfStack = \the\LengthOfStack}
 \ifnum\LengthOfStack=\z@ \Print{Stack is empty}\fi
 \Iteration=\z@
 \loop
 \ifnum\Iteration<\LengthOfStack
  \ShowItem{\Iteration}
  \advance\Iteration \@ne
 \repeat
}

\def\B#1#2{%
\hbox{\vrule\kern-0.4pt\vbox to #2{%
\hrule width #1\vfill\hrule}\kern-0.4pt\vrule}
}


\newif\ifsinglecol   \singlecolfalse

\def\onecolumn{%
  \global\output={\singlecoloutput}%
  \global\hsize=\PageWidth
  \global\vsize=\PageHeight
  \global\ColumnWidth=\hsize
  \global\TextLeading=12pt
  \global\Leading=12
  \global\singlecoltrue
  \global\let\onecolumn=\relax
  \global\let\footnote=\sing@footnote
  \global\let\vfootnote=\sing@vfootnote
  \ninepoint 
  \message{(Single column)}%
}

\def\singlecoloutput{%
  \shipout\vbox{\PageHead\pagebody\PageFoot}%
  \advancepageno
  \ifplate@page
    \shipout\vbox{%
      \sp@pagetrue
      \def\sp@type{plate}%
      \global\plate@pagefalse
      \PageHead\vbox to \PageHeight{\unvbox\plt@box\vfil}\PageFoot%
    }%
    \message{[plate]}%
    \advancepageno
  \fi
  \ifnum\outputpenalty>-\@MM \else\dosupereject\fi%
}

\def\ItemSep{\vskip\ItemSepamount\relax}

\def\ItemSepbreak{\par\ifdim\lastskip<\ItemSepamount
  \removelastskip\penalty-200\ItemSep\fi%
}


\let\@@endinsert=\endinsert 

\def\endinsert{\egroup 
  \if@mid \dimen@\ht\z@ \advance\dimen@\dp\z@ \advance\dimen@12\p@
    \advance\dimen@\pagetotal \advance\dimen@-\pageshrink
    \ifdim\dimen@>\pagegoal\@midfalse\p@gefalse\fi\fi
  \if@mid \ItemSep\box\z@\ItemSepbreak
  \else\insert\topins{\penalty100 
    \splittopskip\z@skip
    \splitmaxdepth\maxdimen \floatingpenalty\z@
    \ifp@ge \dimen@\dp\z@
    \vbox to\vsize{\unvbox\z@\kern-\dimen@}
    \else \box\z@\nobreak\ItemSep\fi}\fi\endgroup%
}


\def\gobbleone#1{}
\def\gobbletwo#1#2{}
\let\footnote=\gobbletwo 
\let\vfootnote=\gobbleone

\def\sing@footnote#1{\let\@sf\empty 
  \ifhmode\edef\@sf{\spacefactor\the\spacefactor}\/\fi
  \hbox{$^{\hbox{\eightpoint #1}}$}\@sf\sing@vfootnote{#1}%
}

\def\sing@vfootnote#1{\insert\footins\bgroup\eightpoint\b@ls{9pt}%
  \interlinepenalty\interfootnotelinepenalty
  \splittopskip\ht\strutbox 
  \splitmaxdepth\dp\strutbox \floatingpenalty\@MM
  \leftskip\z@skip \rightskip\z@skip \spaceskip\z@skip \xspaceskip\z@skip
  \noindent $^{\scriptstyle\hbox{#1}}$\hskip 4pt%
    \footstrut\futurelet\next\fo@t%
}

\def\footnoterule{\kern-3\p@ \hrule height \z@ \kern 3\p@}

\skip\footins=19.5pt plus 12pt minus 1pt
\count\footins=1000
\dimen\footins=\maxdimen


\def\landscape{%
  \global\TEMPDIMEN=\PageWidth
  \global\PageWidth=\PageHeight
  \global\PageHeight=\TEMPDIMEN
  \global\let\landscape=\relax
  \onecolumn
  \message{(landscape)}%
  \raggedbottom
}


\output{%
  \ifLeftCOL
    \global\setbox\LeftBOX=\vbox to \ZoneBSize{\box255\unvbox\ZoneBBOX}%
    \global\LeftCOLfalse
    \MakeRightCol
  \else
    \setbox\RightBOX=\vbox to \ZoneBSize{\box255\unvbox\ZoneBBOX}%
    \setbox\MidBOX=\hbox{\box\LeftBOX\hskip\ColumnGap\box\RightBOX}%
    \setbox\PageBOX=\vbox to \PageHeight{%
      \UnloadZoneA\box\MidBOX\UnloadZoneC}%
    \shipout\vbox{\PageHead\box\PageBOX\PageFoot}%
    \advancepageno
    \ifplate@page
      \shipout\vbox{%
        \sp@pagetrue
        \def\sp@type{plate}%
        \global\plate@pagefalse
        \PageHead\vbox to \PageHeight{\unvbox\plt@box\vfil}\PageFoot%
      }%
      \message{[plate]}%
      \advancepageno
    \fi
    \global\LeftCOLtrue
    \CleanStack
    \MakePage
  \fi
}


\Warn{\start@mess}

\def\mnmacrosloaded{} 

\catcode `\@=12 


\fi


\newcount\fignumber\fignumber=0\relax
\def\fign{\advance\fignumber by 1}

\def\figmod#1
{
	\ifnum#1>0

\input psfig

\def\putfig#1#2#3{\setbox20=\hbox
	{
	\psfig{file=#1,height=#2cm,clip=,angle=#3}
	}
	\centerline{$\vcenter{\box20}$}}

\def\putfigl#1#2#3{\setbox20=\hbox
	{
	\psfig{file=#1,height=#2cm,angle=#3}
	}
	\centerline{$\vcenter{\box20}$}}

	\else

		\input fignot

	\fi
}

\def\em{\ifdim\fontdimen1\font>\z@ \rm\else\it\fi}

\figmod{1}			


\pageoffset{-2.5pc}{0pc}

%
  
%
%
%

\Autonumber  


\pagerange{0-0}    
\pubyear{0000}
\volume{000}

\begintopmatter  

\title{Modelling the number counts of early-type galaxies by pure luminosity 
evolution}
\author{Ping He$^{2}$ and Yuan-Zhong Zhang$^{1,2}$}

\affiliation{$^1$CCAST (World Laboratory), P.O.Box 8730, Beijing 100080, P.R.China}
\smallskip

\affiliation{$^2$Institute of Theoretical Physics, Academia Sinica, P.O.Box 2735, 
Beijing 100080, P.R.China (hemm@itp.ac.cn)}

\shortauthor{P. He et al.}
\shorttitle{Elliptical Galaxy Number Counts}


\acceptedline{ }


\abstract {
In this paper, we explore the plausible luminosity evolution of early-type galaxies
in different cosmological models by constructing a set of pure luminosity evolution
(PLE) models via the choices of the star formation rate (SFR) parameters and 
formation redshift $z_f$ of galaxies, with the observational constraints derived 
from the Hubble Space Telescope (HST) morphological number counts for elliptical and
S0 galaxies of the Medium Deep Survey (MDS) and the Hubble Deep Field (HDF). We find
that the number counts of early-type galaxies can be explained by the pure luminosity
evolution models, without invoking exotic scenarios such as merging or introducing an
additional population. But the evolution should be nearly passive, with a high $z_f$ 
assumed. The conclusion is valid in all of the three cosmological models we adopted in
this paper. We also present the redshift distributions for three bins of observed 
magnitudes in F814w pass-band, to show at which redshift are the objects that dominate
the counts at a given magnitude. The predictions of the redshift distribution of 
$22.5<b_{j}<24.0$ are also presented for comparison with future data.
}

\keywords {
galaxies: elliptical and lenticular, cd - galaxies: evolution - galaxies: luminosity 
function, mass function - cosmology: miscellaneous.
}

\maketitle 

\section{Introduction}

One of the most basic astronomical methods is to simply count the number of galaxies
as a function of apparent magnitude. Such exploration can be traced back to the early
work of Hubble (Hubble 1926), and revived in the seventies after a nearly half 
century's gap (Brown \& Tinsley 1974). Ever since then, it has been widely-used to 
probe the evolutionary history of galaxy populations, or to help estimate the 
parameters of cosmological geometry. It is just these attempts that brought about a 
problem called the excess of faint blue galaxies (FBGs), which remains one of the grand 
astronomical issues for a long time (Koo \& Kron 1992, Ellis 1997). The difficulties 
lie in that one cannot find a logically simple and self-consistent way to explain the
observational data of different aspects. If one tries to reproduce deep blue galaxy 
counts using flat faint-end luminosity functions (LFs), with no evolution (nE) assumed,
the observed number counts show an excess with respect to the expected predictions by a
factor of $\sim$ 4 to 5 at $b_j$ $\sim$ 24, rising up by a factor of $\sim$ 5 to 10 at
$b_j \sim$ 26 (Maddox et al. 1990a; Guiderdoni \& Rocca-Volmerange 1991), and 
continuously increasing to the faintest levels observed at $b_j \sim$ 28. The adoption
of an open geometry can moderate, but still not help ameliorate the situation. The 
introduction of pure luminosity evolution (PLE), which, as a natural consideration,
allows the spectra and luminosity of galaxies to change with redshifts, can provide a
better fit to the faint galaxy number counts in the $b_j$ pass-band, if an open geometry
($q_0 < 0.5$) or a cosmological constant ($\Lambda > 0$) is involved (Broadhurst, Ellis 
\& Shanks 1988; Colless et al. 1990; Cowie, Songaila \& Hu 1991; Colless et al. 1993; 
Fukugita et al. 1990). But as infrared photometric data become available, one finds that
$K$-band (2.2$\mu$m) counts show no excess with respect to the no-evolution predictions 
up to $K \sim 21$ (Cowie et al. 1990; Gardner, Cowie \& Wainscoat 1994), and such PLE
models will overpredict the counts in $K$-band (Cowie 1991). On the other hand, the 
PLE models also overpredict a high-$z$ distribution of galaxies, which is not found in
the results of $z$ surveys of faint galaxies (Broadhurst et al. 1988; Colless et al.
1990; Koo \& Kron 1992 and references therein). To get away from such a dilemma (i.e.,
the optical/infrared and the photometric/spectroscopic paradoxes), a number of less
straitforward models concerning number evolution of galaxies have been proposed. One
is the merger model which would decrease the comoving number density of objects while
increasing their luminosities (Guiderdoni \& Rocca-Volmerange 1991; Broadhurst, Ellis
\& Glazebrook 1992; Carlberg \& Charlot 1992; Kauffmann, Guiderdoni \& White 1994).
Another is to introduce an entirely new population of dwarf galaxies which once existed
in early times and have faded and/or disappeared in recent epochs (Broadhurst et al.
1988; Cowie 1991; Cowie et al. 1991; Babul \& Rees 1992; Babul \& Ferguson 1996).

Another complexity we need to consider is that there may exist some inherent 
uncertainties in the present-day LFs, which is not well-determined by the local 
surveys. For instance, if we adopt a high normalization of the characteristic density,
or consider a steep faint-end-slope, which may accommodate more dwarf galaxies than
the flat one, the degree of excess will be substantially decreased (Saracco, Chincarini
\& Iovino 1996). An extensive investigation of literature shows that there exist many
discrepancies between different research groups (cf. King \& Ellis 1985; Loveday et al.
1992; Marzke et al. 1994a, 1994b; Im et al. 1995; Roche et al. 1996) in the 
determination and/or adoption of the values of the luminosity function (LF) parameters
($\phi^*$, $M^*$, $\alpha$) and the relative mix ratio between different morphological
types. Some authors (De Propris et al. 1995) even announced that an extreme steep 
slope of LF faint-end had been observed, say $\alpha \sim$ -2.2, in the cores of four
rich clusters of galaxies (Abell 2052, Abell 2107, Abell 2199 and Abell 2666). The
uncertainties in local LFs include, e.g., a large local fluctuation ( Shanks 1989),
significant local evolution (Maddox et al. 1990b), selection effects and/or
incompleteness (Zwicky 1957; Disney 1976; Ferguson \& McGaugh 1995), or perhaps
systematic errors in local surveys (Metcalfe, Fong \& Shanks 1995). Thus the
universality of LFs is doubtable, while the well-determination of present-day LF is
of great significance for understanding the galaxy evolution at high redshifts, 
and is conducive to reconciling the paradoxes mentioned previously.
 
Great progress of observational cosmology in recent years has been made through the
powerful ground-based $10-m$ Keck telescope, and especially the Hubble Space Telescope
(HST). The latter, with its high resolution of $0.1''$ FWHM, can provide us with image
information of great value in that the morphologies of different Hubble types can be
segregated into several wide classes (cf. Driver \& Windhorst 1995a; Driver et al 1995b;
Abraham et al. 1996, and references therein). With morphological data, it becomes 
possible to simplify the modelling of FBGs so that each morphological type can be 
allowed to be modelled independently, and hence the complexity of each individual model
can be greatly reduced (Driver \& Windhorst 1995a).

Following this line of thought, we consider first the modelling of E/S0's number counts
in the Medium Deep Survey (MDS) and the Hubble Deep Field (HDF) obtained by WFPC2's
F814w pass-band on board HST in the present paper, by means of PLE models according to
the latest version of population synthesis models of Bruzual \& Charlot (1997, hereafter
BC97),without invoking the more exotic scenarios mentioned above. We choose the
parameters of the star formation rates (SFRs) and the formation redshift ($z_f$) of
galaxies to reproduce the photometric properties such as colors and spectral energy
distributions (SEDs) of the local population. Another motivation of this paper is to 
explore up to what extent the parameters of cosmological models within the framework of
standard cosmology can be constrained by our PLE models. We find that the number 
count-magnitude relation of E/S0 galaxies can be well explained by our PLE model in any
cosmological geometry we adopted in this paper, including a) $\Omega_{0}$=1, $h$=0.5
($H_{0}$=100$h$ km s$^{-1}$ Mpc$^{-1}$), b) $\Omega_{0}$=0.1, $h$=0.5 and c) 
$\Omega_{0}$=0.2, $\lambda_{0}$=0.8, $h=0.6$, under the appropriate choice of the 
parameters of SFRs and $z_f$s. This seems to be conflicting with the result of 
Drive et al. (1996).

In Section 2 we will demonstrate the basic methods for modelling the number counts by
pure luminosity evolution of galaxies and describe the cosmological models as well as
the LF for early-type galaxies adopted in this work. Some details such as dust
extinction are also introduced. The results of our models are shown in Section 3 for
comparison with the observational data. In Section 4, we will discuss the influence of uncertainties in LF, dust extinction and different initial mass function (IMF) upon our results. A comparison of our work with others is also made in this section. We will give our summary and conclusions in Section 5.

\section{Modelling The Number Counts of Ellipticals}

The number distribution of galaxies between the interval $(m,z) \sim (m+dm,z+dz)$ is
stated by the expression as follows: $$d^2N(m_{\lambda},z)={\omega \over 4\pi} {dV 
\over dz}\phi(M_\lambda) dm_{\lambda} dz, \eqno(1)$$ where $\omega$ is solid angle,
$dV$ is the comoving volume element of the redshift interval $z \sim z+dz$, and $\phi$
is the present-day LF. The relation between apparent magnitude $m$ and $M$ in the 
$\lambda$-pass band is given by $$ m_{\lambda}=M{_\lambda}+5log({d_{L}\over 10})+corr,
\eqno(2)$$ where $d_{L}$ is the luminosity distance measured in $pc$ and dependent on
cosmology. The term $corr$ refers to the correction needed to translate the galaxy 
observer-frame magnitude into its rest-frame magnitude. In the nE model, it is only
characterized by $k$-correction, which accounts for the red-shifting of the spectra
due to the recession of galaxies. The $e$-correction should be considered if the
intrinsic galaxy luminosity evolution is involved, and in more realistic cases, some
secondary effects, e.g., dust extinction, should also be taken into account to some
extent. The magnitude-limited number counts ($z$-distribution) can be obtained by
integrating (1) over the specific magnitude range $(m_{1}, m_{2})$. By integrating
over redshift $z$ the differential number counts (number-magnitude relation) of
galaxies can be obtained. Since galaxies will become drastically faint beyond the
redshift $z_L$ at which the Lyman continuum break is shifted into the effective 
wavelength of the pass-band being considered (Madau 1995), the upper limit of the
integral over $z$ should be $z_{up}$=min$(z_f,z_L)$, where $z_f$ is the assumed $z$
of galaxy formation. For $B$-band $z_{L}$ is 4.0, while for F814w it is 7.8. We will
see the significance of this consideration in what follows.

\begintable{1}
\nofloat
\caption{{\bf Table 1.} Parameters of the cosmological models under consideration,
as well as the assumed formation redshifts of galaxies.}
\halign{%
\hfil#\hfil & \hfil#\hfil & ~~\hfil#\hfil & ~~\hfil#\hfil & ~~\hfil#\hfil~ \cr
\noalign{\vskip 3pt}\noalign{\hrule}\cr
Scenario~~&~~~$h^a$~~~&~~~$\Omega_0$~~~&~~$\lambda_0$~~&~~${z_{f}}^b$~~\cr
\noalign{\vskip 3pt}\noalign{\hrule}\cr\noalign{\vskip 3pt}
A  & 0.5   &  1.0  &   0    &  3.5 , 5 , 8 , 10    \cr
B  & 0.5   &  0.1  &   0    &  2.5 , 3.5 , 5 , 10  \cr
C  & 0.6   &  0.2  &   0.8  &  2.5 , 3.5 , 5 , 10  \cr
\noalign{\vskip 3pt}\noalign{\hrule}\cr
}
\tabletext{\noindent
$^a$ $H_0 = 100 h$ km s$^{-1}$ Mpc$^{-1}$;\par\noindent
$^b$ $z_f$ refers to the redshift of galaxy formation.
}
\endtable

\subsection{Cosmological Models}

As one of the goals of our present exploration, we would examine the validity of our
PLE models under the three choices of the currently popular cosmological models. The
first we consider is $\Omega_{0}$=1 and $H_{0}$=50km s$^{-1}$ Mpc$^{-1}$, a flat
Einstein-de Sitter universe favoured by the standard inflation theory. The second we
adopted is the Friedmann-Robertson-Walker model, in which we adopt $\Omega_{0}$=0.1,
$H_{0}$=50 km s$^{-1}$ Mpc$^{-1}$, representing the low density universe. The third
is $\Lambda$-dominated Friedmann-Lema\^{i}tre model for which we choose 
$\lambda_{0}$=0.8, $\Omega_{0}$=0.2 and a higher Hubble constant $H_{0}$=60km s$^{-1}$
Mpc$^{-1}$, also compatible with the inflation theory. Both the low density and the 
$\Lambda$-dominated universes have larger volumes at a given redshift. We list the
parameters of these cosmological models in Table 1. Also listed in it are the galaxy
formation redshifts we assumed in the three cosmological models.

Once a specific cosmological model is given, the relations between redshifts,
look-back times and volume elements, which are needed in modelling, can be determined
completely by the standard formulae (cf. Weinberg 1972; Guiderdoni \& Rocca-Volmerange
1990; Fukugita et al. 1990). 

%
\begintable{2}
\nofloat
\caption{{\bf Table 2.} Observed colors of local elliptical and S0 galaxies 
for reference.
}
\halign{%
#\hfil~~~~~~~~~~~~~~~~~  & ~~\hfil#\hfil~~~\cr
\noalign{\vskip 3pt}\noalign{\hrule}\cr\noalign{\vskip 3pt}
~~~~~~~CI & observed \cr
\noalign{\vskip 3pt}\noalign{\hrule}\cr\noalign{\vskip 3pt}
~~~~~${U-B^a}$       & 0.43   \cr
~~~~~${B-V^a}$       & 0.95   \cr
~~~~~${B-R^b}$       & 1.83   \cr
~~~~~${b_j-I^c}$     & 2.39   \cr
~~~~~${b_j-K^b}$     & 4.16   \cr
\noalign{\vskip 6pt}\noalign{\hrule}\cr
}
\tabletext{\noindent
 $^a$ Mixed from Fukugita et al. (1995);
\par\noindent $^b$ Mixed from Yoshii et al. (1988);
\par\noindent $^c$ Mixed from Yoshii et al. (1988) and Lidman et al. (1996).
}
\endtable

\subsection{Evolutionary SEDs of Galaxies}

There are standard evolutionary synthesis techniques (Bruz-ual \& Kron 1980; Guiderdoni
\& Rocca-Volmerange 1987; Charlot \& Bruzual 1991; Bruzual \& Charlot 1993, hereafter
BC93; Bruzual \& Charlot 1997, hereafter BC97) to obtain the SEDs of galaxies. Our
galaxy SEDs are computed on the basis of the latest galaxy isochrone synthesis spectral
evolution library (GISSEL96, BC97). The BC97 models are built from a library of stellar
tracks in the theoretical H-R diagram, covering all evolutionary stages of stars with
multi-metallicity. The spectra of galactic stars, from near-UV to near-IR, extending to
the far-UV by means of model atmospheres, are employed in the synthesis. For the present 
purpose, however, we do not model the evolution of metallicity with respect to $z$, and 
consider the solar metallicity only.

Colors of galaxies, as a significant photometric property, can provide us with valuable
information about their present-day composition of stars, and accordingly give us
important clues to the formation and evolution of galaxies. In the present work, we use
several broad-band colors of elliptical galaxies (see Table 2) observed locally as the 
preliminary constraints on our selections of the model parameters. The photometric
systems we use here are: Johnson's $UBVRI$ systems (Johnson \& Morgan 1953), Couch's
$b_{j}$ and $r_{f}$ (Couch \& Newell 1980) and the Palomar 200 IR detectors' $K$ band
filter (effective wavelength, 2.2$\mu$m). We set the photometric zero point by the SEDs
of the star of $\alpha$ Lyre. Thus we can compute the model colors to make comparison
with the observed ones, and among them we select the SFR parameters to give the best
fit.

Throughout this work, we adopt the Scalo (1986) IMF for our
models. The Scalo IMF is less rich in massive stars than the Salpeter (1955) IMF
because of the steeper slope of the former at the high-mass end. By the adoption of
Scalo IMF, the UV flux of early times can be greatly reduced so as to avoid a large
number of galaxies being detected at high $z$, which are not observed in current deep
surveys (Pozzetti, Bruzual \& Zamorani 1996, hereafter PBZ96).  

%
\fign\beginfigure{\fignumber}
\putfigl{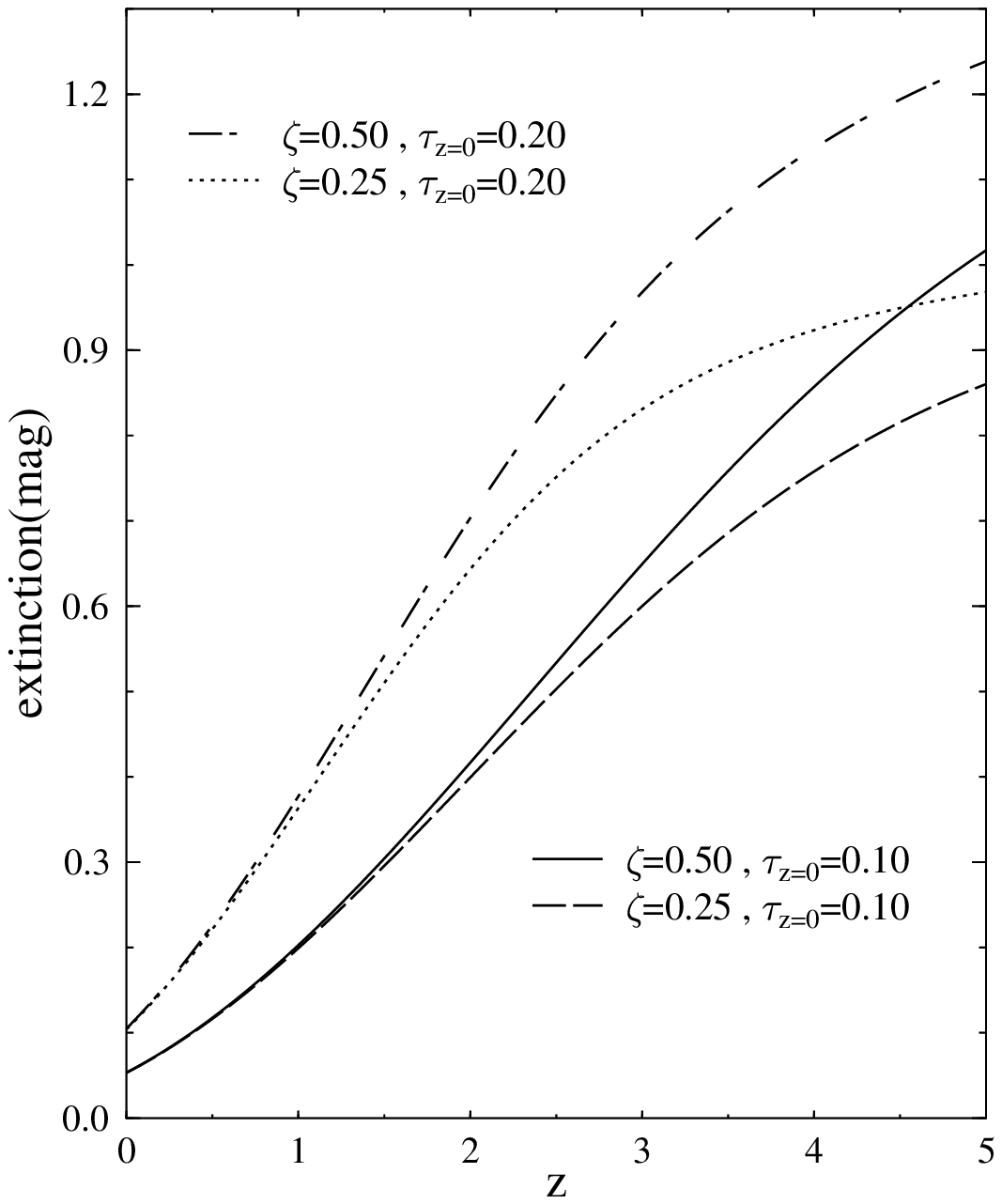}{10}{0}
{
{\bf Figure 1.}
Extinction in the $B$-band as a function of redshift for an $L^*$ galaxy. The predictions of models are shown by lines, with the model parameters (see Wang 1991 for denotations of these parameters) exhibited in the figure.
}
\endfigure

\subsection{Dust Extinction}

Simple luminosity evolution models (e.g., PBZ96) only consider the contributions to the
fluxes by star populations. But in reality, most galaxies are full of interstellar
medium (ISM), and hence effects of internal absorption by dust are also needed to be
introduced into the modelling. In particular, the dust extinction has great influence
upon $z$-distributions of galaxies (Wang 1991; Gronwall \& Koo 1995; Campos \& Shanks
1995).

Unlike the situation in spiral galaxies, however, physical and evolutionary relationships between the various components of the ISM in ellipticals are not yet well understood (Goudfrooij, 1996). To circumvent the complexity and uncertainty in determining the law of extinction for ellipticals, we make an {\it ad hoc} assumption that Wang's prescription, which is just the simulation for spirals, could be extrapolated to the description of dust extinction for ellipticals, except that we choose a larger geometrical parameter $\zeta$=0.50, and a smaller optical depth $\tau_{z=0}=0.10$, in $B$-band for present-day $L^*$ ellipticals, which mimic the geometrical feature and less dust content for local ellipticals, in contrast with the adoption by Wang, that $\zeta=0.25$ and $\tau_{z=0}=0.20$ for spirals. We also assume that the optical depth depends on the galaxy luminosity as $\tau \propto L^{0.5}_{z=0}$, which is a simple representation of the fact that luminous galaxies are seen to be much dustier than those of lower luminosity. The extinction curve of our models is taken to be a power law in wavelength, namely $\propto \lambda^{-n}$, with n = 2 (Draine \& Lee 1984), which is the same as Wang's adoption. By such choices, the extinctions for $B$-band are 0.05 mag at present-day, and 1.02 mag at $z=5$; for F814w they are 0.02 and 0.50 at $z=0$ and 5, respectively. Observationally, extinction corrections in $B$ band for ellipticals in the nearby 4 clusters are 0.18mag (A2256, $z$=0.0601), 0.07mag (A2029, $z$=0.077), 0.05mag (Coma, $z$=0.023), and 0.04mag (A957, $z$=0.045), obtained from the NASA Extragalactic Database (cf. Schade, Barrientos \& L\'opez-Cruz 1997). Hence, the results of our models can be accepted when compared with the observation. We show the relation of extinction against redshift in $B$-band in Figure 1. It can be seen that present-day extinctions are not affected by the geometrical parameter $\zeta$, but are sensitive to the local optical depth $\tau_{z=0}$; while $\zeta$ will play an important role at high redshifts, at which fluxes are greatly reduced, and hence galaxies become much fainter by a thick geometrical parameter than the thin one.

Needless to say, it is only a simple treatment, since we do not model the dust evolution
and some other details such as the inclination of galaxies are also neglected. But even
so, the effect of dust extinction can greatly reduce the UV flux in early epochs due to
the wavelength dependence on the extinction law. 

\begintable*{3}
\nofloat
\caption{{\bf Table 3.} $z_f$ and SFR parameters listed in column 2 and
3 in the three cosmological models under consideration, with the modelled 
colors listed from column 4 to 9.
}
\halign{%
~#\hfil~&~~~\hfil#\hfil~~&~~~\hfil#\hfil~&\hfil#\hfil&\hfil#\hfil&
\hfil#\hfil~&\hfil#\hfil~&\hfil#\hfil~&\hfil#\hfil\cr
\noalign{\vskip 5pt}\cr 
%
\noalign{\hrule}\cr
\noalign{\vskip 1pt}
 & & & & Scenario A & & & & \cr
\noalign{\vskip 2pt}
~Models&z$_f$&$\tau^{a}_{e}$&~~$U-B$~~~&~~~$B-V$~~~&~~~$B-R$~~~&
~~~$b_{j}-I$~~~&~~~$b_{j}-K$~~&~~~$b_{j}-I_{F814}$~~ \cr
\noalign{\vskip 5pt}
\noalign{\hrule}\cr
\noalign{\vskip 1pt}
~~A1   &  3.5 & 0.8 & 0.64 & 0.92 & 1.67 & 2.27 & 4.10 & 2.03 \cr
~~A2   &   5  & 0.2 & 0.65 & 0.92 & 1.69 & 2.30 & 4.15 & 2.05 \cr
~~A3   &   8  & 1.0 & 0.64 & 0.92 & 1.68 & 2.29 & 4.13 & 2.04 \cr
~~A4   &  10  & 1.0 & 0.65 & 0.92 & 1.68 & 2.29 & 4.13 & 2.04 \cr
\noalign{\vskip 5pt}
\noalign{\hrule}\cr
\noalign{\vskip 16pt}
%
\noalign{\hrule}\cr
\noalign{\vskip 1pt}
 & & & &Scenario B & & & & \cr
\noalign{\vskip 2pt}
~Models&z$_f$&$\tau^{a}_{e}$&~~$U-B$~~~&~~~$B-V$~~~&~~~$B-R$~~~&
~~~$b_{j}-I$~~~&~~~$b_{j}-K$~~&~~~$b_{j}-I_{F814}$~~ \cr
\noalign{\vskip 5pt}
\noalign{\hrule}\cr
\noalign{\vskip 1pt}
~~~B1   &  2.5 & 0.2 & 0.67 & 0.93 & 1.70 & 2.32 & 4.19  & 2.07 \cr
~~~B2   &  3.5 & 0.1 & 0.69 & 0.94 & 1.72 & 2.34 & 4.22  & 2.09 \cr
~~~B3   &   5  &  0  & 0.71 & 0.95 & 1.74 & 2.36 & 4.25  & 2.11 \cr
~~~B4   &  10  & 1.0 & 0.71 & 0.95 & 1.74 & 2.36 & 4.25  & 2.11 \cr
\noalign{\vskip 5pt}
\noalign{\hrule}\cr
\noalign{\vskip 16pt}
%
\noalign{\hrule}\cr
\noalign{\vskip 1pt}
 & & & &Scenario C & & & & \cr
\noalign{\vskip 2pt}
~Models&z$_f$&$\tau^{a}_{e}$&~~$U-B$~~~&~~~$B-V$~~~&~~~$B-R$~~~&
~~~$b_{j}-I$~~~&~~~$b_{j}-K$~~&~~~$b_{j}-I_{F814}$~~ \cr
\noalign{\vskip 5pt}
\noalign{\hrule}\cr
\noalign{\vskip 1pt}
~~~C1   &  2.5 & 0.4  & 0.70 & 0.93 & 1.72 & 2.33 & 4.20  & 2.07 \cr
~~~C2   &  3.5 & 0.2  & 0.70 & 0.94 & 1.73 & 2.35 & 4.23  & 2.09 \cr
~~~C3   &   5  & 0.05 & 0.72 & 0.95 & 1.74 & 2.37 & 4.26  & 2.11 \cr
~~~C4   &  10  & 1.2  & 0.71 & 0.95 & 1.74 & 2.36 & 4.25  & 2.11 \cr
\noalign{\vskip 5pt}
\noalign{\hrule}\cr
}
\tabletext{\noindent
$^a$ $\tau_{e}$ is the characteristic time-scale for SFRs, measured in Gyr,
and $\tau_{e} = 0$ represents the single burst model, in which the evolution
is pure passive.
}
\endtable

\subsection{Luminosity Function}

The present-day luminosity function of galaxies is well represented by the Schechter
(1976) analytic form as follows
$$\phi(L)dL=\phi^*{\left({L\over L^{*}}\right)^{\alpha}}e^{-L/L^{*}}d\left({L\over L^*} 
\right), \eqno(3)$$ where $L^*$ is the characteristic luminosity and $\alpha$ is the
faint-end slope. $\phi^*$ is the characteristic density, which is the normalization
related to the number of luminous galaxies per unit volume (cf. Ellis 1997). 

We adopt the morphology-dependent LF from the models of Roche et al. (1996,
hereafter RSMF96) for our present study. For early-type galaxies, the parameters of LF
assumed in RSMF96 are $M_{B}^{*}=-21.00$, $\alpha=-0.70$ and $\phi^{*}=9.68 \times 10^{-4}$Mpc$^{-3}$. We derive the local LF of $I_{F814}$ band by shifting $M_{B}^{*}$ according to the present-day color $B-I_{F814}$ computed by our models (see the last column in Table 3). Note that in RSMF96, the E/S0 galaxies are divided into two types, namely, the cold and the hot ellipticals. In our case, however, we do not make such a distinction and simply incorporate them into a single type . Besides, it should be mentioned that the
classification between early-type, late-type spirals and irregulars of RSMF96 LF is
also suitable for future investigations, although we only deal with ellipticals in
the present work. For PLE models, in contrast to the cases of number evolutions, the 
evolution of LF with respect to $z$ is realized only by $e$-corrections ( with the dust
extinction involved), which are functions of $z$.

Considering that there exist some uncertainties in determining the local LF, as
mentioned in the introduction, we attempt to vary the parameters of LF, in particular,
$\alpha$, to show to what extent such uncertainties can affect our conclusions.

%
\fign\beginfigure*{\fignumber}
\putfigl{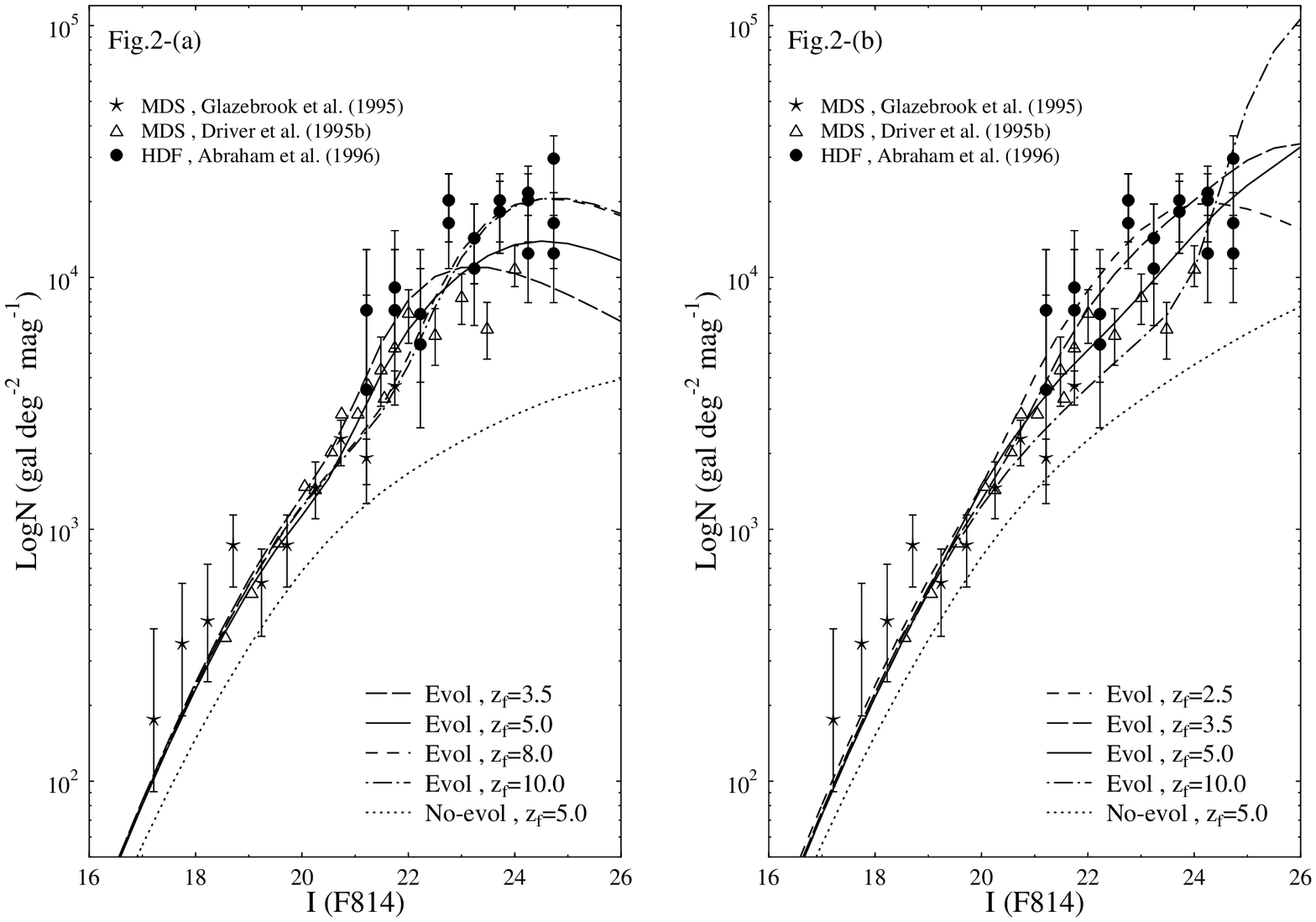}{9.9}{0}
\vskip 2pt
\caption
{
{\bf Figure 2.}
The differential number counts for E/S0 galaxies as a function of apparent
magnitude in $I_{F814}$ band. Figures 2-(a), (b) and (c) are for Scenario A,
B, and C, respectively. The sources of the observational data are exhibited
in the figure and these data are indicated by symbols. The models are shown
by lines.
}
\endfigure

%
\fign\beginfigure{\fignumber}
\putfigl{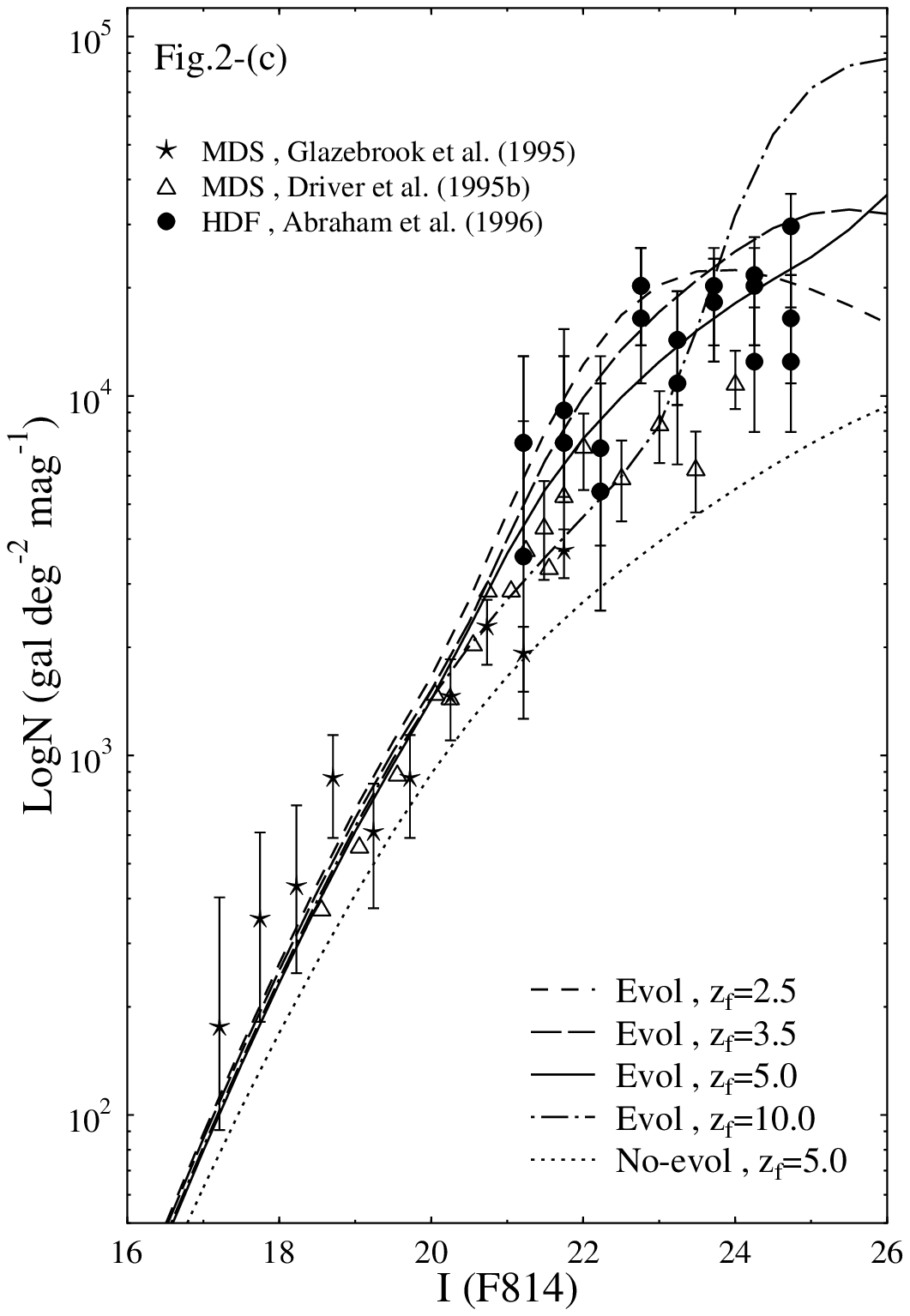}{10}{0}
\endfigure

\section{Results}
\subsection{Star Formation Rates}

As outlined previously, we adjust the model parameters of SFRs to fit best the
observational local colors of E/S0 galaxies. The SFR is a single-parameter function
taking the form of exponential decrease with respect to time, i.e., $\psi(t) \propto
exp(-t/\tau_{e})$, which is a natural result under the simple assumption that $\psi(t)$
is proportional to the available gas (Kennicutt 1983), where $\tau_{e}$ is the e-folding
time characterizing this form of SFRs. Thus in the viewpoint of modelling, the
evolutionary photometric and spectroscopic properties of galaxies can be completely
determined by these parameters in our simple PLE models once the cosmological models are
specified. We summarize the values of these model parameters in Table 3 in column 2 and
3. The model colors of $U-B$, $B-V$, $B-R$, $b_{j}-I$ and $b_{j}-K$ are listed from
column 4 to 8, for comparison with the observational ones shown in Table 2. The 
$b_{j}-I_{F814}$ colors are also given in column 9 for converting the LF from $b_j$ into
$I_{F814}$ band.

Table 3 shows that the modelled colors in a specific band computed by any one of our
models are much close to each other, the difference from model to model is very small.
The model colors are also close to the observed ones listed in Table 2. In particular,
the BC97 models can reproduce better the $b_{j}-K$ color than the old version (BC93) by
about $0.2-0.3$mag reddening. Considering that the uncertainties in local colors of
galaxies are about $0.1-0.2$mag, the adopted values of the parameter $\tau_{e}$ should
be acceptable. Furthermore, we can also find that the models can grossly reproduce the
observed SEDs for E/S0 galaxies of present-day by comparing the $k$-corrections computed
from the models with the empirical one (not shown). 

Superior to the old version as it is, the BC97 models still need further refinement,
especially in the UV range, for the modelled $U-B$ colors are slightly redder than the
observed one, which signals such improvements to be needed.

\subsection{HST F814w Number Counts of Ellipticals}

The wider spatial coverage of the MDS and the improved signal-to-noise of the HDF,
which provide us with abundant image information, now allow us to model the
morphologically segregated galaxy number counts, and we pay attention to the number
counts of E/S0 galaxy in our current investigation. The observed data of elliptical
galaxy number counts are taken from Glazebrook et al. (1995) and Driver et al. (1995b)
for MDS, and Abraham et al. (1996) for HDF. Although there may exist some errors in the
morphological classification (the scheme and details in classification have been
discussed in the above-quoted works), from Figure 2 we can see that the data show gross
agreement within the overlapped regions between different research group's, indicating,
to some extent, that the classification is reliable. Another noteworthy feature is the
flattening beyond $I_{F814} \sim 22.5$, which is just the unique property that is
different from the counts of other types (spirals or irregulars) as well as the overall
population.

Figures 2-(a), (b) and (c) show differential number counts derived from the F814w 
bandpass on board HST, together with our models in the three scenarios A, B and C indicated by lines. Apparently, the non-evolutionary predictions cannot reproduce the data, either at faint magnitudes or at the bright end. The predictions with luminosity evolution, 
however, can match better the number counts at bright end up to $I_{F814} \sim 21.0$mag,
no matter which scenario, and which $z_f$ in each scenario is taken into account, i.e.,
the number counts at bright magnitudes are insensitive to geometry or $z_f$. The effect
of luminosity evolution is significant and should not be neglected. Furthermore,
allowing for the fluctuation of the Glazebrook et al. sample at bright magnitudes, we
believe that the normalization of the RSMF96 LF employed in this work is reasonably 
well.

Discrepancies between these evolutionary predictions do exist and occur beyond 
$I_{F814}>21.0$mag. We will discuss them in detail the the following.

\subsubsection{Scenario A}

All the predictions fall within the error bars of the data, while only the model of
$z_f=5$ predicts better the number count than the others. Besides, it can grossly
reproduce the faint-end flattening at around $I_{F814} \sim 23.0-25.0$. For the case of
$z_f=3.5$, though it keeps on increasing up to $\sim 23.0$mag and reproduces better the
counts even than that of $z_f=5$ model at $I_{F814} \sim 21.0-23.0$, it will turnover
consequently and greatly underpredict the faint-end counts as well as the flattening.
It seems that $z_f=3.5$ for a flat Einstein-de Sitter world model is not proper, since
it is too low, and hence there is not enough volume to accommodate sufficient objects.
While if the $z_f$ is chosen as high as 8 or 10, the models will predict much lower
counts from $I_{F814}$=20.5 to 22.5 than the others.

\subsubsection{Scenario B and C}

Models of $z_{f}=10$ can be absolutely ruled out: the predictions at $I_{F814}>24.0$
overshoot the HDF data by about a factor of 3 to 4, indicating that volumes in the two
world models at such a redshift are both too large; they simultaneously underpredict
the counts at $I_{F814}=21.0 - 24.0$ mag (Scenario B) or 21.0 - 23.0 mag (scenario C),
indicating that luminosity evolutions in the two cases are both too low.

The prediction seems better for $z_f=2.5$ in Scenario B, since it can reproduce well
the counts almost at every magnitude from 21.0 to 25.0 mag. It is also not bad for
Scenario C, though the prediction at 23.0 - 24.0 mag is slightly higher than that of
Scenario B. 

The predictions are not as good as those of $z_{f}=2.5$ for models of $z_{f}=5.0$ in
either Scenario B or C. In particular, the slopes $\gamma \equiv dlogN(m)/dm$ of the
models at $I_{F814} > 22.5$ mag are $\sim$0.22 and $\sim$0.18 for Scenario B and C,
respectively, steeper than the flattening of the observational data. It, however, should
be attributed not to the evolution but to the geometry, since the evolution is nearly
passive (see Table 3). As can be seen, the slopes even for non-evolutionary models are
still as high as $\sim$0.12 and $\sim$0.14 for Scenario B and C, respectively.
Regardless of the steep slopes, there is neither sufficient nor convinced evidence to
exclude such models as $z_f=5.0$. After all, our models reproduce the counts within the
magnitude range where there exist observational data ($I_{F814}=17.5 - 25.0$ mag).

The $z_{f}=3.5$ models of Scenario B and C can be regarded as interpolations between
the models for $z_f=2.5$ and $z_f=5$, but they are more similar to those of $z_f$=5.

%
\fign\beginfigure*{\fignumber}
\putfigl{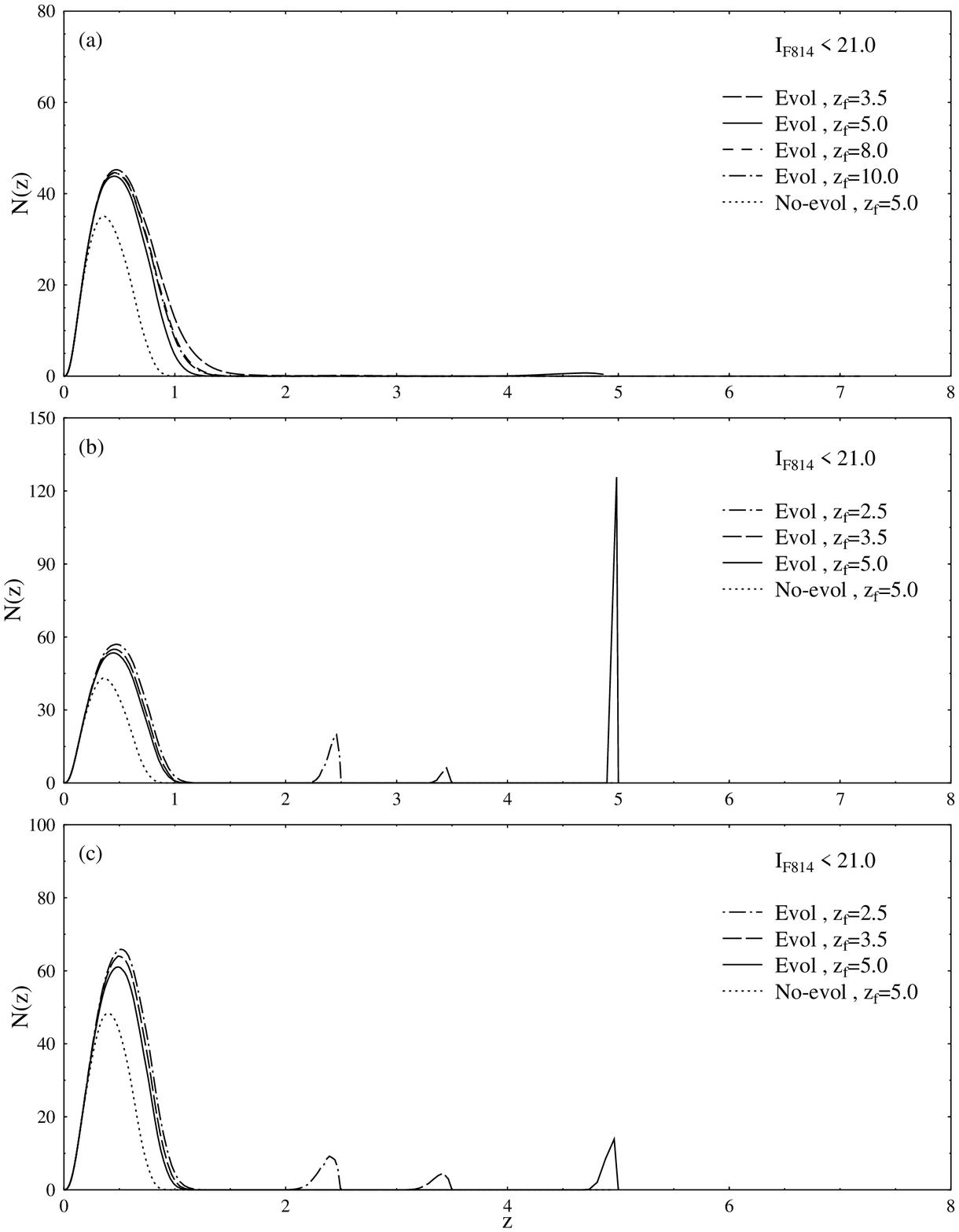}{18}{0}
\vskip 2pt
\caption
{
{\bf Figure 3.}
The $z$-distribution of F814w $<21.0$. Panels (a), (b) and c are for Scenario 
A, B and C respectively. Models are indicated by lines. The model predictions 
have been normalized in the unit of deg$^{-2}$ with redshift bin $\Delta z$=0.01.
}
\endfigure

%
\fign\beginfigure*{\fignumber}
\putfigl{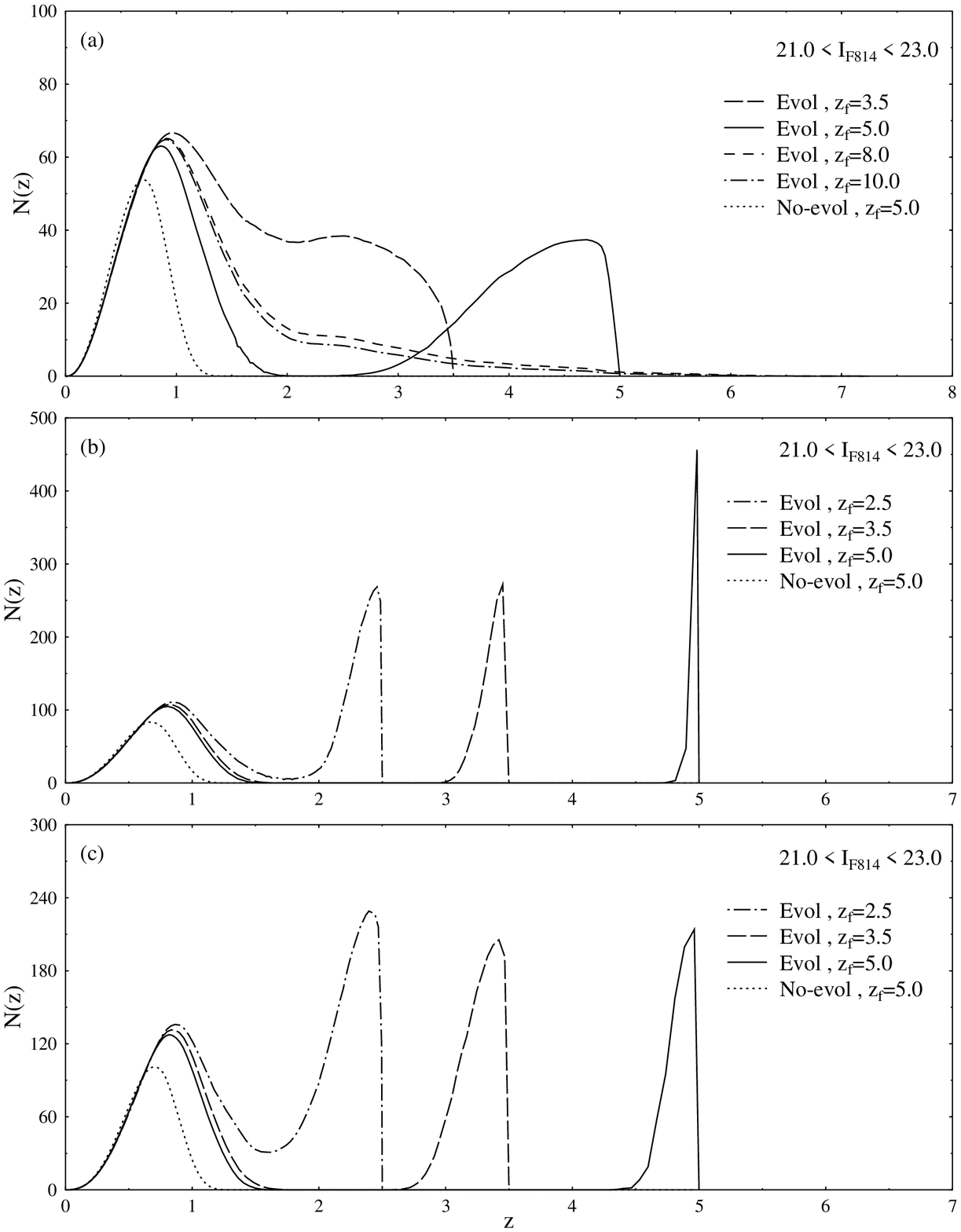}{18}{0}
\vskip 2pt
\caption
{
{\bf Figure 4.}
The $z$-distribution of $21.0<$ F814w $<23.0$. Panels (a), (b) and c are for Scenario
A, B and C respectively. Models are indicated by lines. The normalization is the same
as Figure 3.
}
\endfigure

%
\fign\beginfigure*{\fignumber}
\putfigl{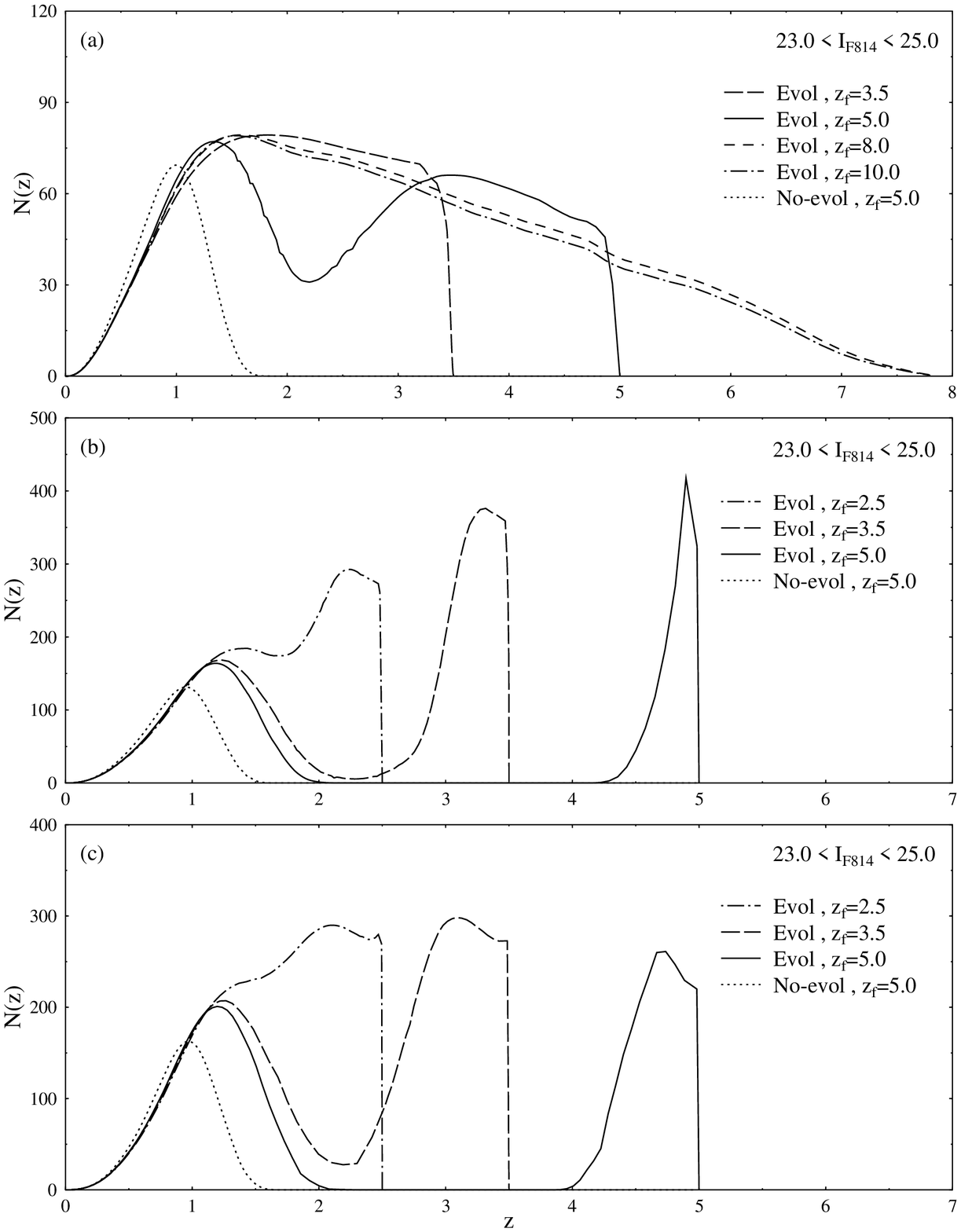}{18}{0}
\vskip 2pt
\caption
{
{\bf Figure 5.}
The $z$-distribution of $23.0 <$ F814w $<25.0$. Panels (a), (b) and c are for Scenario
A, B and C respectively. Models are indicated by lines. The normalization is the same
as Figure 3 and Figure 4.
}
\endfigure

%
\fign\beginfigure*{\fignumber}
\putfigl{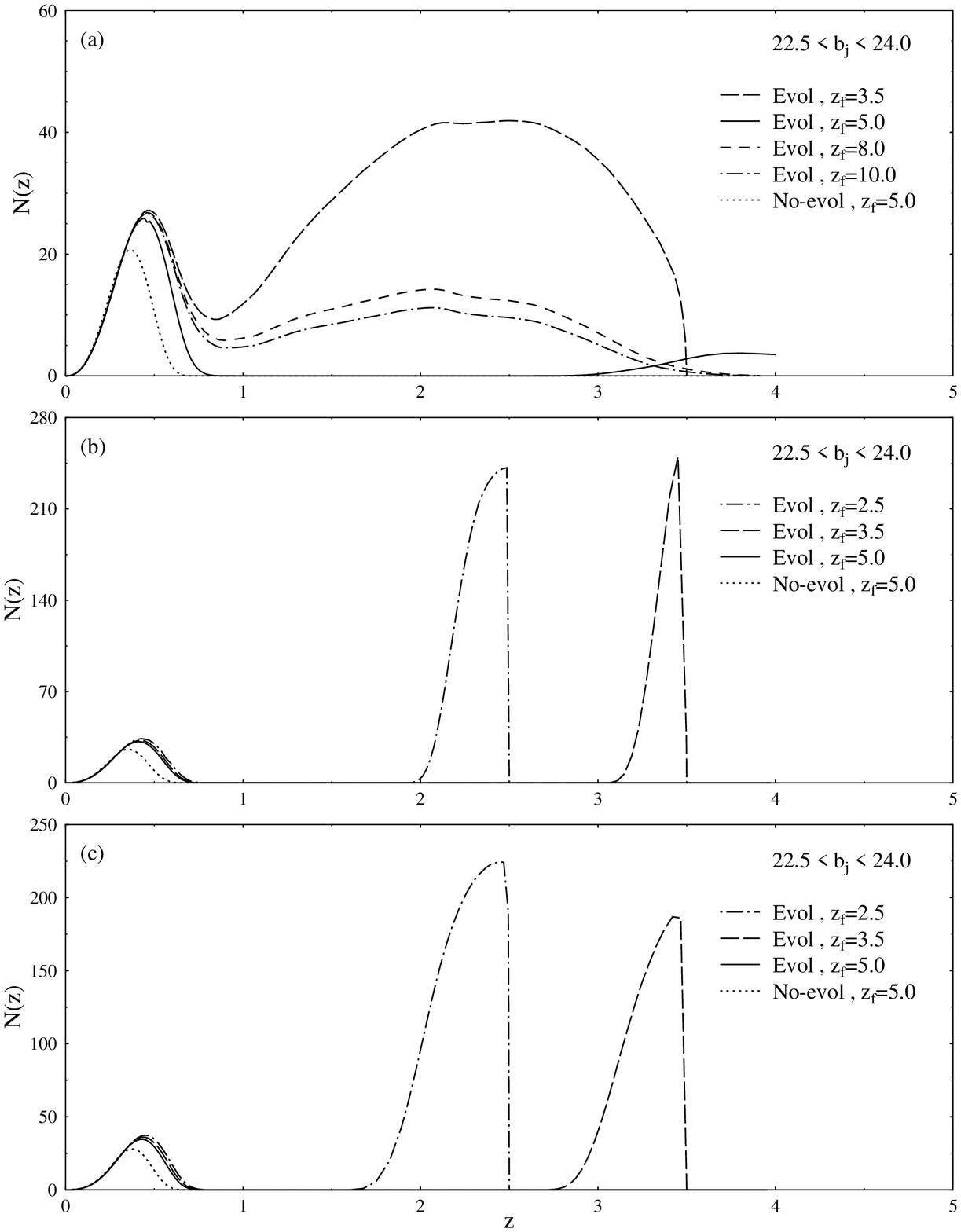}{18}{0}
\vskip 2pt
\caption
{
{\bf Figure 6.}
The $z$-distribution of $22.5 < b_{j} < 24.0$. Panels (a), (b) and (c) are for Scenario
A, B and C respectively. Models are indicated by lines. The normalization is the same
as Figure 3 to Figure 5.}
\endfigure

%
\fign\beginfigure*{\fignumber}
\putfigl{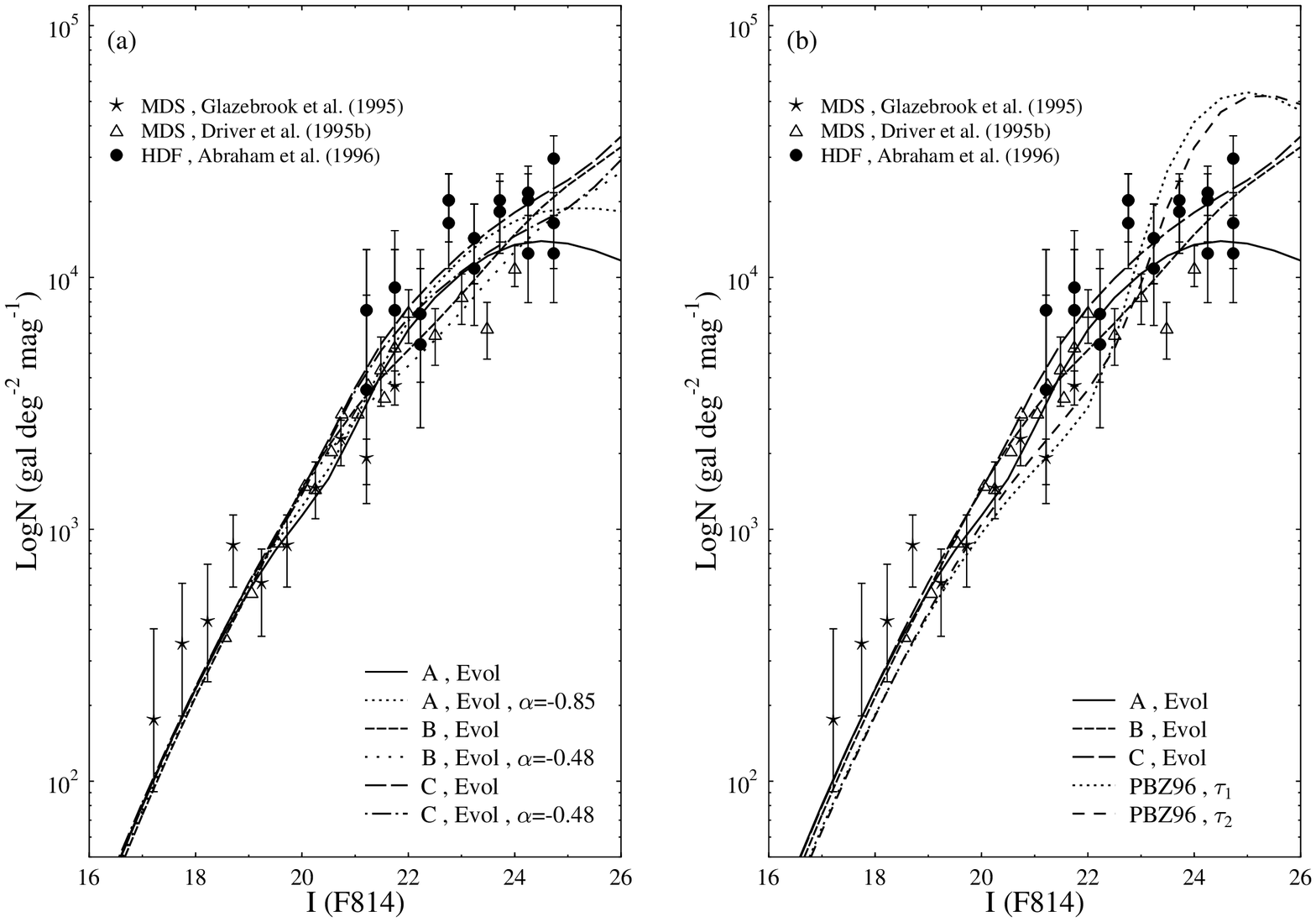}{9.9}{0}
\vskip 2pt
\caption
{
{\bf Figure 7.}
The differential number counts for E/S0 galaxies as a function of apparent magnitude in
$I_{F814}$ band. The models are for $z_f=5.0$, which are shown by lines. The letter A,
B and C in the figure represent Scenario A, B and C, respectively. Panel (a): 
Comparison with the results of varying the faint-end slope of LF; (b): Comparison with
the results of PBZ96.
}
\endfigure

\subsection{Redshift Distribution}

In modern astronomy, redshift distributions of galaxies are the key measurement required
to probe the local luminosity function, or possible evolution taking place in galaxies
as well as the geometry of the universe. Such measurement has already reached up-to-date 
as faint as $B \sim 24$ mag by the new spectroscopic samples derived from the Keck 
Telescope (Cowie et al. 1995, 1996). It can be seen from Section 3.2, that the
conclusions are rather uncertain and obscure. Hence, we should incorporate redshift
distributions for bins of observed magnitudes, to continue our investigation. Such
exploration will be conducive to the understanding of our models, though such
spectroscopic data has not been obtained morphologically as yet.

\subsubsection{z- distribution of $I_{F814}$-band}

(a) $I_{F814}<21.0$

In Figure 3 we plot our predictions for the redshift distributions of $I_{F814}<21.0$
mag. Panels (a), (b) and c are for Scenario A, B and C, respectively. We re-scale our
predictions in the unit of deg$^{-2}$ with redshift bin $\Delta z$=0.01. We can see
that all the models with evolution will peak at $z\sim0.5$ (hereafter, we call it the
first), but it can also be seen that there is another peak existing at the redshift of
$z_f$ (hereafter, we call it the second), except for Scenario A. In particular, the
second peak of $z_{f}=5.0$ model in Scenario B is more than two times as high as the
first. An extremely rapid declining of the SFR (for the $z_f=5.0$ model in Scenario B,
the time scale $\tau_{e}=0$, i.e., single burst) will consume (almost) all the gas in
galaxies to form stars (almost) in an instantaneous burst, rendering the galaxies to
be detectable at that high redshift, not bringing about the subsequent star formation,
especially of massive stars, and resulting in the rapid fading in luminosity of galaxies
shortly after the formation of stars of the first generation. Therefore, the existence
of the second peak should be completely attributed to the rapid luminosity evolution.
Yet the peak is very narrow, and hence it will not be expected to contribute too much
to the counts.

\vskip 5pt
\noindent (b) 21.0$<I_{F814}<23.0$

Figure 4 is plotted for the predictions of the redshift distribution limited in 
21.0$<I_{F814}<23.0$, the others are the same as Figure 3. For Scenario B and C, it
can be seen that the high-$z$ peaks of the $z_f=5.0$ models are even higher than those
in the case of $I_{F814}<21.0$, but the contributions to the counts are still not
comparable to the first. As for Scenario A, it is obvious that the high-$z$ contribution
begins to be significant for the models whose $z_{f}$=3.5 or 5. What forms the sharp
contrast is that there are no such peaks in the predictions of the $z_f$=8 or 10 models,
It is just the absence of these peaks that leads to the lower number counts at around
$I_{F814}\sim20.5$ to $\sim$22.5 mag than those of $z_f$=3.5 or 5.0 models (see Figure
2-(a)). 

\vskip 5pt
\noindent (c) 23.0$<I_{F814}<25.0$

Figure 5 is for the redshift distributions predicted by our models within the magnitude
range 23.0$<I_{F814}<25.0$. All the curves of the models are similar to those in the
case of 21.0$<I_{F814}<23.0$, except that high-$z$ contributions become dominant in
such faint magnitude range, especially for Scenario A, as can be seen from the figure.

In each of the three figures above-mentioned, we have also plotted predictions of
non-evolutionary models of $z_{f}=5.0$ for each panel to make comparison with the
evolutionary predictions. The most significant difference between the non- and 
the evolutionary models can be seen from these figures is that no high-$z$ peaks appear
in the predictions of non-evolutionary models. Hence, according to the PLE models, and
differing from nE ones, the objects at high-$z$ also contribute to the predicted number
counts at magnitudes fainter than $I_{F814}\sim 20.5$. Our PLE models presented here 
are characterized by these high-$z$ peaks in the predictions of redshift distributions.   

It is worthwhile to mention that these peaks, in principle, are detectable through the
F814w band-pass since $z_L$ for F814w is $\sim7.8$, which is larger than the redshift
at which these peaks are seen.

\subsubsection{$z$- distribution of $22.5<b_{j}<24.0$}

Since there are both abundant and deep enough spectroscopic samples in $B$-band surveys
up to date, the predictions of $z$ distributions in this passband is of great
significance. In Figure 6, we present the predicted $z$ distributions by our PLE models
within the magnitude range $22.5 < b_{j} < 24.0$. From  panels (a), (b) and (c) we can 
see that: 1) The high-$z$ peaks for $z_f=5.0$ models are absent in all of the three
cosmological models adopted currently. The shapes of these curves are similar to those
of non-evolutionary ones, except that for Scenario A, where there is merely a little
rise beyond $z$=3, and 2) the high-$z$ peaks or tails still exist in $z_f$=2.5 or 3.5
models in any scenario. As mentioned in Section 2, the light coming from beyond $z_L$
will be greatly reduced by Lyman absorption, hence the objects beyond $z_L$ are hardly
detectable, and that $z_f>4.0$ (for b-band, $z_L=4.0$) is favoured by the $z$
distribution of $22.5<b_{j}<24.0$ in any world models. On the contrary, $z_f$=2.5 or
3.5 are not appropriate choices, unless the high-$z$ tail or the high-$z$ peak can be
detected in the future.

%
\fign\beginfigure*{\fignumber}
\putfigl{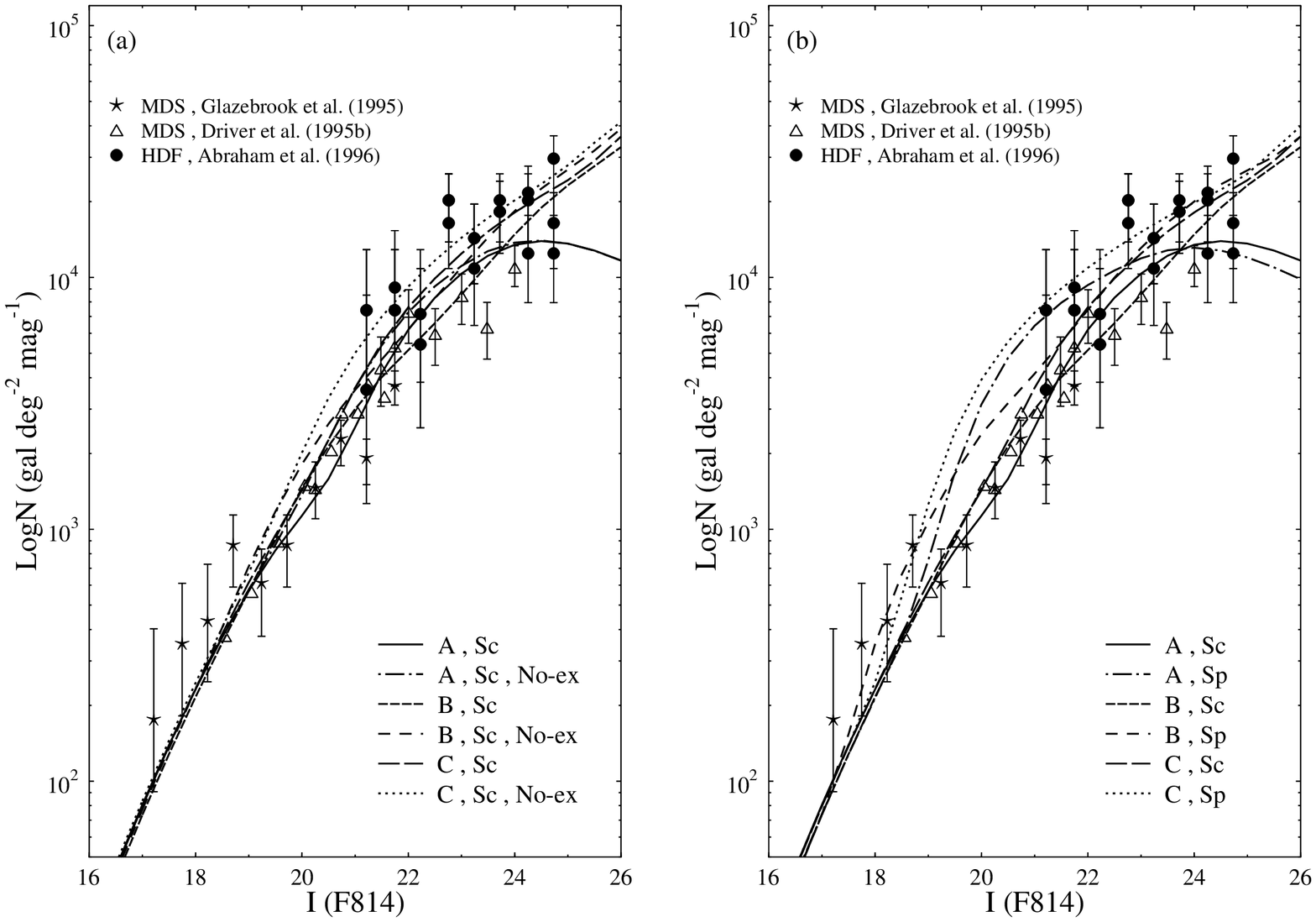}{9.9}{0}
\vskip 2pt
\caption
{
{\bf Figure 8.}
The differential number counts for E/S0 galaxies as a function of apparent magnitude in $I_{F814}$ band. The models are for $z_f=5.0$, which are shown by lines. In the two panels, the symbols `Sc' and `Sp' denote Scalo and Salpeter IMF, respectively. `No-ex' means no dust extinction is involved. Panel (a): Comparison with the results of no dust extinction; (b): Comparison with the results by assuming Salpeter IMF.
}
\endfigure

%
\fign\beginfigure*{\fignumber}
\putfigl{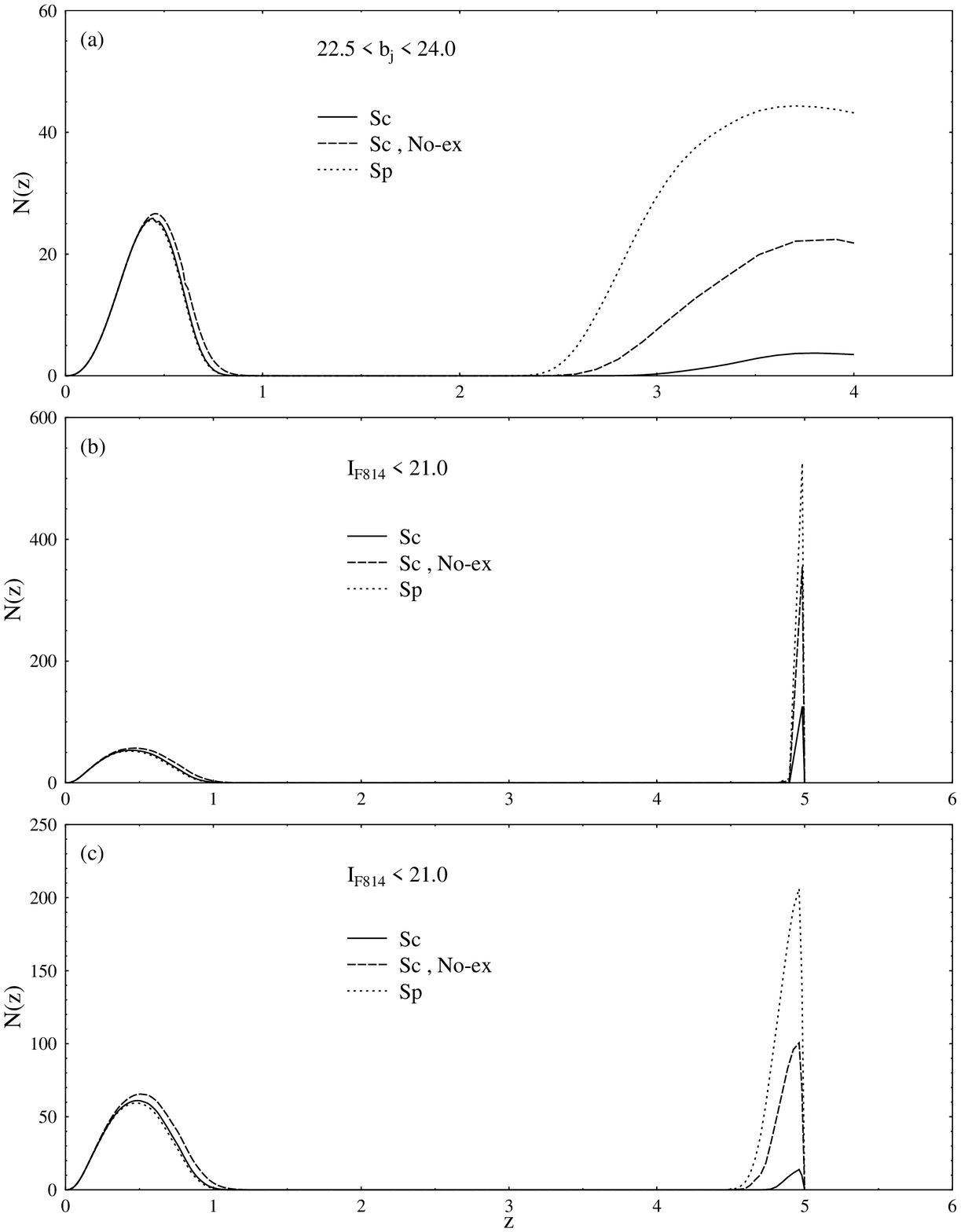}{18}{0}
\vskip 2pt
\caption
{
{\bf Figure 9.}
The $z$-distribution for ellipticals. Models are indicated by lines. Meanings of the symbols are the same as those in Figure 8. Panel (a) is for Scenario A, limited in $22.5<b_j<24.0$; panels (b) and (c) are for Scenario B and C respectively, limited in $I_{814}<21.0$. 
}
\endfigure

\section{Discussion}

\subsection{Uncertainties in LF}

As mentioned in Introduction, there may exist many uncertainties in determining the
local LF, hence there is no reason in believing that the LF in our present work is the
unique adoption. We will explore  the influence of the uncertainties in LF upon our
conclusions, in particular, upon the number counts, simply by varying the LF parameters.
From the previous investigation, we have already seen that the normalization, which
involves both $\phi^*$ and $M^*$ (cf. Ellis 1997), is appropriate. Thus we only
concentrate on the faint-end slope $\alpha$. The following concerns only the $z_f$=5.0
models.

For Scenario A, we choose a higher value for $\alpha$, say, -0.85, which is slightly
lower than -0.9, the value of Marzke et al. (1994a, 1994b). For Scenario B and C, we
adopt -0.48, the value of PBZ96. From Figure 7-(a) we can see that such variation will
not affect the counts at bright magnitudes, and will not bring about too much change at
the faint-end, either. For Scenario A, such variation leads to only 0.12dex higher than
that of $\alpha=-0.70$ at $I_{F814}=25.0$. But the faint-end slope of the number-count
curve $\gamma$ is better than its counterpart of $\alpha=-0.70$. As for Scenario B and
C, such variation does not make too much difference. Hence, through such prescriptions,
we have verified that the $uncertainties$ in LF (will) not influence too much our 
conclusions.

\subsection{Comparison with Other Works}

We notice that the PLE model by PBZ96, which was constructed under the BC93 spectral
evolution models in a world model with $\Omega\sim 0$, can account for most of the
observed photometric and spectroscopic properties of galaxies, including the number
counts in the $U, b_{j}, r_{f}, I$ and $K$ bands, as well as the color and redshift 
distributions derived from most of the existing samples. We have interest in examining
whether the parameters of the PBZ96 model adopted for E/S0 are suitable for the
modelling of early-type galaxy number counts in the $I_{F814}$ passband. The $e+k$-
correction for E/S0 galaxies in $I_{F814}$ band is computed by employing BC93 models,
exactly in terms of the parameters adopted by PBZ96. From Figure 7-(b), we can see that,
compared with our models, neither the $\tau_1$ nor the $\tau_2$ model of PBZ96
(Following PBZ96, $\tau_1$ and $\tau_2$ refer to the time scale of SFR $\tau_{e}$ being
1Gyr and 2Gyr, respectively, with IMF adopted as Scalo one for ellipticals.) can
reproduce the number counts of E/S0 galaxies properly. Furthermore, the adoption of a
$\Omega \sim 0$ cosmological model is obviously unphysical.

As can be seen from Section 3.2, our PLE models can reproduce the number counts well
in all three cosmological models under consideration, especially for Scenario C (the
world model dominated by a $\lambda$), as seems to be conflicting with the result of
Driver et al. (1996), who concluded that flat models dominated by a cosmological
constant are ruled out from comparison of their E/S0 counts (Driver et al. 1995b)
with their model predictions. The difference between our models and Driver et al.'s
(1996) lies in the normalization of E/S0 galaxy counts, i.e., a much higher
normalization adopted by Driver et al. Besides, the E/S0 counts for MDS of Driver
et al. are apparently lower than the HDF counts of Abraham et al. (1996).

\subsection{Dust extinction and IMF}

We have taken into account to some extent the influence of dust extinction on the predictions of the models, as described in Section 2.3. From Figure 8-(a), we can see that the effect of dust extinction is obvious, especially for Scenario B and C, such that models without extinction will predict more ellipticals than observed from $I_{814}\sim 19$ to $\sim 21$. Moreover, the high-$z$ tails or peaks are even higher than the results if dust is not considered in the models (see Figure 9).

The Scalo IMF is more favoured by the observations than the Salpeter IMF. It can be seen from Figure 8-(b), even with dust extinction involved, the models can not reproduce satisfactorily the number counts, and greatly overestimate the counts between $I_{814}\sim 19 -21$. A less steep IMF at high-mass end such as the Salpeter one will lead to more massive stars existing at early times, rendering UV fluxes are so strong that more galaxies can be detected at high-$z$ (see Figure 9).

\section{Summary and Conclusions}
\def\item#1{\item{\rm (#1)}}

We believe that in any realistic evolutionary models, the luminosity evolution must be
considered, since there exists a well-known fact that galaxies are composed  of stars,
and stars continuously come into birth and evolve into the post-main-sequence. The 
lifetime of stars depends on their mass. This leads to the continuous change of the
photometric and spectroscopic properties of galaxies. Although nE models can work
sometimes (Driver \& Windhorst 1995a), it is obviously unphysical and can only be
treated as baseline for comparison with observations and other models. In this paper,
we have constructed a series of PLE models in the three cosmological models to explain
the observed number counts for elliptical galaxies in MDS and HDF obtained by the HST.
We summarize our investigations in the following.

\beginlist

\item{i)} Although some of our model colors do not match quite well the observed ones,
for example, in all cases, the $U-B$ color is slightly redder than the observed, the
success of the BC97 population synthesis model which our present work is based upon is
still worthy of being affirmed. More recently, Steidel et al. (1996) found that
high-redshift galaxies are also well described by the (BC93) models. It should be
pointed out that some complexities are neglected in the current prescription of simple
PLE models, e.g., we have not modelled the evolution of metallicity in our work. There
exists, however, a so-called {\it age-metallicity degeneracy} stating that for stellar
populations older than 2Gyr, an increase in metallicity by a factor of 2 and a decrease
in age by a factor of 3 results in almost identical optical and near-IR colors (Worthey
1994). Therefore, when a formation redshift $z_{f}$ is assumed so that the age of 
galaxies is determined in a specific cosmological model, the effect of evolution of
metallicity can be partly compensated by the appropriate adoption of the parameters of
SFR $\tau_{e}$. This will also from the other side indicate that the local colors are
insensitive to the adoption of $z_{f}$, as can also be seen from Table 3. 

\item{ii)} The Hubble constant $H_{0}$ plays just a minor role in our present 
investigation. Although $H_{0}$ is included in the normalization $\phi^*$ and 
characteristic luminosity $L^*$ of LF, it can be canceled out by the quantities 
relative to cosmology. The only aspect needed to pay attention to is that the age of
galaxies is measured in $H_{0}^{-1}$. Hence the age of galaxies will be smaller if a
larger $H_{0}$ is adopted, and vice versa, as will more or less influence the modelled
spectra of galaxies. The adoption of $H_{0}$=50km s$^{-1}$ Mpc$^{-1}$ seems quite
reasonable for a flat world model with $\Omega=1$. We can see from Table 3 that the
integrated colors computed by the latest BC97 spectral evolutionary models are
appropriate compared with the observed locally (Table 2). Hence regardless of the
$H_0$/globular cluster age problem, the flat cosmological model will not suffer from
the conflict of color/age in the context of BC97 models.

\item{iii)} In spite of the fact that the models of $z_f=5.0$ failed to predict the
faint-end flattening of the number counts in Scenario B or C, our PLE models can still 
be regarded to have reproduced well the data from the MDS and the HDF up to $I_{F814}
\sim 25$ within the range of uncertainties in all the three scenarios. The corresponding
predicted $z$-distribution of $22.5<b_{j}<24.0$, though we have not observed data to
compare with, does not present a high-$z$ tail or peak. The shapes of evolutionary
predictions are similar to those of no-evolution for the $z$ distribution, with the mean
redshift $z_{m}$ slightly higher than the no-evolution's, as can be attributed to the
effect of luminosity evolution.

\item{iv)} The above conclusions are not affected by the variation of the LF parameters
($\phi^*$, $\alpha$ and $L^{*}$). Hence, even though there may exist some uncertainties
and biases in the determination of local LF, the above conclusions are still valid and
reliable.

\item{v)} The law of dust extinction for ellipticals is largely uncertain. We simply extrapolate the Wang's approach to the current case, and the results seem to be well in agreement with the observed extinction in $B$ band for present-day ellipticals. By such an {\it ad hoc} assumption, together with the adoption of Scalo IMF for ellipticals, UV fluxes can be greatly reduced at high redshifts, leading to the high-$z$ peaks substantially depressed. Moreover, the predictions of number counts are also improved.

\item{vi)} High $z_f$, say 5.0, and nearly-passive evolution shortly after a (nearly) 
single burst to form most of the stars (in our work, $\tau_e$=0.2, 0 and 0.05Gyr for 
Scenario A, B and C, respectively) are critical for the modelling, as is just the case
of the conventional scenario for formation and evolution of ellipticals (cf. Eggen, Lynden-Bell \& Sandage 1962; Partridge \& Peebles 1967). On the contrary, it will not be favoured by, especially, spectroscopic samples that galaxies formed at as low as $z_f=2.5$ (or 3.5), or the evolution is as large as that of PBZ96, unless more exotic scenarios are introduced into the modelling.

\item{vii)} We do not touch upon the possibilities of number evolutions in the present 
investigation, though it has been widely considered to account for the optical/infrared
and the photometric/spectroscopic paradox (see Introduction). Even though there may be
the possibility that a new population of dwarf galaxies existed at high-$z$ and then 
faded or disappeared lately, it is unlikely that they are of early-type galaxies, or at
least most of them can not be early-types. If so, the faint-end slope of number counts
of early-type galaxies will be steeper than the currently observed data from HST. The
assumption of mergers is also not necessary here, though the possibility of mergers is
not ruled out. If the scenario of merging is taken into account, a series of problems
will be involved (cf. T\'{o}th Ostriker 1992; Babul \& Ferguson 1996, and references 
therein), and the physical mechanism for mergers is much more complicated than that of 
PLE.

\item{viii)} We have chosen three cosmological models for our work. The predictions of
number counts by the flat Einstein-de Sitter model agree well with the observed data,
so do the open Friedmann-Robertson-Walker model ($\Omega_{0}$=0.1, $\lambda_{0}$=0, 
$H_{0}$=50 km s$^{-1}$ Mpc$^{-1}$) and the Friedmann-Lema\^{i}tre model
($\Omega_{0}$=0.2, $\lambda_{0}$=0.8, $H_{0}$=60 km s$^{-1}$ Mpc$^{-1}$). Therefore, 
we can not discriminate between these cosmological models by the galaxy number counts. 
Our conclusion here contradicts with that of Driver et al. (1996), who announced that 
the world models dominated by a $\lambda$ are ruled out by their modelling of E/S0 
number counts. Like many other authors (Tinsley 1972; Ellis 1997), we believe that the
formation and evolution of galaxies is a problem far from being well understood, and 
any definite judgment about world models, whether the universe is open, flat, closed or
$\Lambda$-dominated, can not be made as yet by the number counts.

\item{ix)} As mentioned previously, our PLE models do not include the evolution of 
metallicity with respect to $z$, and those details such as the recycling of the residual
gas ejected by the dying stars as well as some selection effects are also ignored in our
models. These can be considered as further improvement over our present models.

\endlist

All in all, the models presented in this work can explain well the number counts of 
elliptical and S0 galaxies derived from HST under the assumption of PLE. It is unlikely 
that the problem of FBGs is caused by early-type galaxies, and we hold identical views 
as other researchers (Driver \& Windhorst 1995; Driver et al. 1995; Glazebrook et al. 
1995; Abraham et al. 1996) on this. Furthermore, our work shows that the cosmological
parameters can not be determined by up-to-date observations. In a forthcoming paper, we
will devote ourselves to the galaxy number counts of the other morphological types 
segregated by HST, namely, the early-type spirals (Sabc) and late-type spirals and 
irregulars (Sdm/Irr), as well as the overall populations, with the present results and 
conclusions involved, which will be examined further.

\section*{Acknowledgments}

P. He should like to thank Mr. S. Charlot for providing us with their synthesis spectral
evolutionary models and helpful correspondences. We acknowledge the valuable discussion 
between Dr. B.F. Roukema, Prof. Z.G. Deng and Prof. Z.L. Zou. The authors are much 
grateful to Mrs. B. Poggianti for her constructive comments and suggestions to improve the manuscript of this paper. We also thank The State Key Laboratory of Science and Engineering Computing (LSEC) of Academia Sinica for providing us with computer supports. This work is in part supported by the National Natural Science Foundation of China.


\bye